\documentclass[longbibliography,aps, prx, amsmath, amssymb, amsfonts, twocolumn,superscriptaddress]{revtex4-2}
\usepackage[german,american]{babel}
\usepackage{graphicx}
\usepackage{dcolumn}
\usepackage{bm}
\usepackage{amssymb}
\usepackage{amsmath}
\usepackage{amsfonts}
\usepackage{epsfig}
\usepackage{mathbbol}
\usepackage{latexsym}
\usepackage{dcolumn}
\usepackage{subfigure}
\usepackage{hyperref}
\usepackage{float}
\usepackage[normalem]{ulem}
\usepackage[usenames, dvipsnames]{color}
\usepackage{epstopdf}
\usepackage{comment}

\hypersetup{                    
	colorlinks,
	citecolor=blue,
	filecolor=blue,
	linkcolor=blue,
	urlcolor=blue
}
\usepackage[utf8]{inputenc}

\begin{document}
	\title{Quantum Approximate Optimization Algorithm with Adaptive Bias Fields}
	
	\author{Yunlong Yu}
	\affiliation{ State Key Laboratory of Low Dimensional Quantum Physics, Department of Physics,\\Tsinghua University, Beijing 100084, China}
	\affiliation{Kavli Institute for Theoretical Sciences, University of Chinese Academy of Sciences, Beijing, China}
	
	\author{Chenfeng Cao}
	\affiliation{Department of Physics, The Hong Kong University of Science and Technology,\\ Clear Water Bay, Kowloon, Hong Kong, China}
	
	\author{Carter Dewey}
	\affiliation{Department of Physics, University of Wisconsin–Madison, 1150 University Avenue, Madison, Wisconsin 53706, USA}
	
	\author{Xiang-Bin Wang}
	\affiliation{ State Key Laboratory of Low Dimensional Quantum Physics, Department of Physics,\\Tsinghua University, Beijing 100084, China}
	
	\author{Nic Shannon}
	\affiliation{Theory of Quantum Matter Unit, 
		Okinawa Institute of Science and Technology Graduate University, Onna-son, Okinawa 904-0412, Japan}
	
	\author{Robert Joynt}
	\affiliation{Department of Physics, University of Wisconsin–Madison, 1150 University Avenue, Madison, Wisconsin 53706, USA}
	\affiliation{Kavli Institute for Theoretical Sciences, University of Chinese Academy of Sciences, Beijing, China}

	\begin{abstract}
		The quantum approximate optimization algorithm (QAOA) transforms a simple many-qubit wavefunction into one which encodes a solution to a difficult classical optimization problem.  It does this by optimizing the schedule according to which two unitary operators are alternately applied to the qubits.  In this paper, the QAOA is modified by updating the operators themselves to include local fields, using information from the measured wavefunction at the end of one iteration step to improve the operators at later steps. It is shown by numerical simulation on MaxCut problems that, for a fixed accuracy, this procedure decreases the runtime of QAOA very substantially.  This improvement appears to increase with the problem size. Our method requires essentially the same number of quantum gates per optimization step as the standard QAOA, and no additional measurements. This modified algorithm enhances the prospects for quantum advantage for certain optimization problems.
	\end{abstract}
	\date{\today}
	\maketitle
	\section{Introduction}
	We are in the era of Noisy Intermediate-Scale Quantum (NISQ) devices~\cite{preskill2018quantum}. This motivates the development of variational quantum algorithms (VQA) that use a sequence of relatively short quantum circuits with parameters that are iteratively updated by a classical optimizer~\cite{peruzzo2014variational,cerezo2020variational2, bharti2021noisy}. 
	VQAs have been designed for a wide range of problems, such as ground state and excited state preparation~\cite{kandala2017hardware, wang2019accelerated, nakanishi2019subspace, zeng2020variational,cao2021energy},
	quantum state diagonalization~\cite{larose2019variational, cerezo2020variational}, quantum data compression~\cite{romero2017quantum, bondarenko2020quantum, cao2021noise}, quantum fidelity estimation~\cite{cerezo2020variationalf, chen2020variational}, and quantum compiling~\cite{sharma2020noise}. 
	
	The Quantum Approximate Optimization Algorithm (QAOA) is the leading example of a VQA for combinatorial optimization\cite{farhi}. The repeated quantum evolution depends on classical parameters
	which are iteratively updated. The final result is a calculated value for the cost function and a corresponding quantum state that encodes an approximate solution to a classical optimization problem. The QAOA is considered to be a good candidate for an algorithm that will be superior to classical algorithms reasonably soon \cite{harrow}, so many studies have focused on the experimental demonstrations for the QAOA in different physical systems~\cite{harrigan2021quantum,otterbach2017unsupervised,qiang2018large,pagano2020quantum,willsch2020benchmarking,abrams2019implementation,bengtsson2020improved}. However, it is generally thought that the standard QAOA will not be competitive with established classical methods until a time when quantum machines are considerably larger than they are today \cite {guerreschi, moll}. Thus there is intense activity to improve the QAOA~\cite{farhi2017quantum,yang2017optimizing,brady2021optimal,wecker2016training,li2020quantum,bravyi2020obstacles,lukin,zhu2020adaptive}, which would bring this time closer. That is also the goal of the present work. 

	The QAOA starts the quantum computer in the ground state of the mixing Hamiltonian $H_M^\mathrm{s}$ and then alternately applies the unitary operators  $\exp(-i\gamma_k H_C)$ and $\exp(-i\beta_k H_M^\mathrm{s})$, where $H_C$ is the problem Hamiltonian whose ground state is sought~\cite{farhi}. At level $p$, $\{\gamma_k\}_{k=1}^{p}$ and $\{\beta_k\}_{k=1}^{p}$ are two sets of parameters that fix the schedule of the evolution. At each iteration, $\{\gamma_k\}_{k=1}^{p}$ and $\{\beta_k\}_{k=1}^{p}$ are improved by measuring $ H_C $. (Henceforth we drop the subscripts and superscripts on $\{\gamma_k\}_{k=1}^{p}$ and the other parameter sets.)  
	
	Simulations on classical computers have shown some impressive results for the QAOA as applied to MaxCut \cite {crooks, lukin}. The authors of Ref.~\cite{lukin} produced an efficient iterative scheme that runs in time $O$(poly $p$) and that approached the known solutions with high accuracy.  They demonstrated the superiority of QAOA over standard quantum annealing - the classical optimization effectively isolates the small gap events that plague annealers and the Quantum Adiabatic Algorithm and substantially neutralizes them, though it should be noted that modifications of quantum annealing can do this for certain special problems \cite{muthukrishnan,ozguler}    
	
	Here we introduce a method which can greatly accelerate convergence of the QAOA, through the use of adaptive bias fields in the mixing Hamiltonian (ab-QAOA). Varying the ansatz operator in the VQA~\cite{grimsley,tang2021qubit} or QAOA ~\cite{hadfield,zhu2020adaptive}  has been proposed before, but the ab-QAOA approach has two critical differences to earlier protocols. The first is that previous modifications of the QAOA do not use all the information available at the end of a step.  One can use the energy measurement in more than one way to guide the system toward its ground state.  The second is that local fields are introduced in $H_M^\mathrm{ab}$ as was done previously in quantum annealing \cite {grass} and the starting state is reinitialized accordingly. Some methods such as FALQON~\cite{kenny1,kenny2} also use measurements to update the operators, but the operators applied are completely different.  
	
	Mean field theory is the usual starting point for the investigation of ordered spin systems.  It can also be very useful for certain Ising spin glass systems, the case of interest here. The prime example is the Parisi solution of the Sherrington-Kirkpatrick model \cite{young}.  Hence it is natural to include elements of mean field theory in any search for a ground state.  Thus the overall philosophy of our approach is to make a marriage between mean field theory and an iterative variational procedure. There are 4 reasons to suppose this will improve the QAOA. 
	
	\begin{enumerate}
	
	\item The optimization is guided not only the energy but also by the local magnetization, so additional information available from the measurements is used. 
	\item Mean field theory is often the best starting point for a variational calculation on a system with many degrees of freedom. 
	\item For any algorithm based fundamentally on the adiabatic theorem, the shorter the distance in Hilbert space from the starting wavefunction final correct ground state, the better the chances of success~\cite{grass}.  Our procedure includes a modification of the wavefunction at each stage of the iteration. 
	\item For any problem whose solution is one of the computational basis vectors (an Ising problem, in condensed matter theory language) a local field term in the $z$ direction will steer the solution in a good candidate direction, owing to the fact that the solution lies in the set of ground states of some local-field Hamiltonian.  
	\end{enumerate}

	Bias fields have previously been introduced in quantum adiabatic algorithm to improve accuracy (defined below)~\cite{grass}, but in that reference the procedure was not adaptive.  This leads us to call our method the ``adaptive bias QAOA''  (ab--QAOA).  The use of adaptive bias fields improves both the accuracy of the solution  and its fidelity, i.e. the overlap between the computed final state and the actual target state. There have been some adaptive QAOA methods, such as the operator pool method ~\cite{zhu2020adaptive}, in that the mixing Hamiltonian is updated, but  re-initialization have not been employed in the past. 
	
	This paper is organized as follows. In Sec.~\ref{sec:algorithm}, we give a detailed description of the MaxCut problem, the QAOA and the ab-QAOA. In Sec.~\ref{sec:result}, the relative performances of the QAOA and the ab-QAOA are computed and analyzed. We also investigate in detail the effects of the bias fields in the ab-QAOA. The conclusion of the paper is given in Sec.~\ref{sec:conclusion}. 
	
	\section{algorithm details}\label{sec:algorithm}
	
	\subsection{MaxCut problem}
	The performance of a heuristic algorithm must be judged against competitors.  In what follows we compare the ab-QAOA, against the standard QAOA (henceforth referred to simply as QAOA) described above. The QAOA has already been compared to classical algorithms \cite{crooks}, so this way of proceeding indirectly benchmarks the QAOA against classical competitors as well. Following Ref.~\cite{farhi} we define the accuracy as 
	\begin{equation}
		r = \frac{E_{\mathrm{opt}}(\psi_{f})}{E_{\max}(\psi_{\max})} \; ,
		\label{eq:r}
	\end{equation}
	where $E_{\mathrm{opt}}(\psi_{f})$ is the expectation value of the problem Hamiltonian in the state $\psi_{f}$ produced by the algorithm and $E_{\max}(\psi_{\max})$ is the value in the optimum state $\psi_{\max}$.
	
	The problem we use for benchmarking ab--QAOA is MaxCut, a canonical problem in graph theory~\cite{mezard2009information}. Let an undirected graph be denoted by $G(V,E)$, where $V$ is the $n$-vertex set and $E$ is the edge set.  The edges may or may not be assigned weights.  If they are, then the weights are chosen uniformly at random from the interval $[0,1]$.  In the unweighted version we wish to partition $V$ into two subsets $V_1$, $V_2$ in such a way as to make the number of edges connecting $V_1$ and $V_2$ as large as possible.  In the weighted version, the total weight of the partition is maximized.  

	\begin{figure}[ht]
		\centering
		
		\includegraphics[scale=0.5]{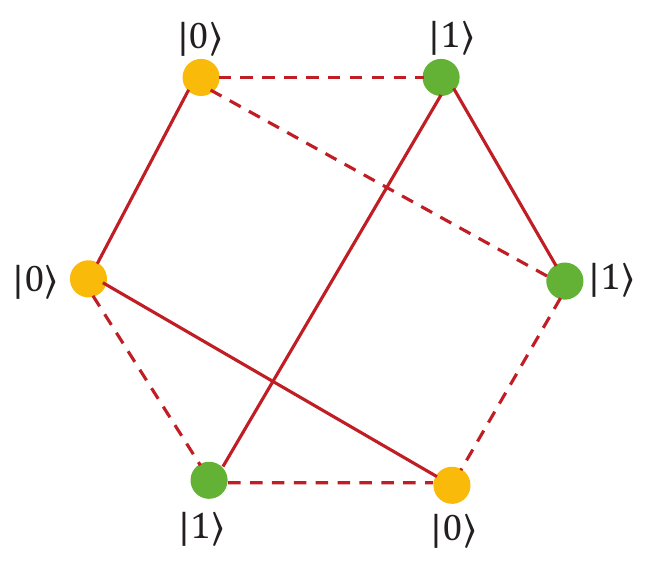}
		\caption{MaxCut problem on an unweighted $3$-regular $6$-vertex graph. Different colors give the different states $|0\rangle$ and $|1\rangle$, and represent the two different subsets $V_1$ and $V_2$ of the vertex set. The problem is to find the division of the vertices that maximizes the number of edges connecting the two subsets.  The dashed edges in the figure represent the cut in this case.} \label{fig:maxcut}
	\end{figure}

	We convert MaxCut to an $n$-vertex Ising model as follows. Define a Pauli matrix $Z_j$ to act on the $j$th vertex and use the eigenstates $|0\rangle$, $|1\rangle$ of the $Z_j$ to represent  $V_1$ and $V_2$.  Thus, in operator language, the MaxCut problem Hamiltonian for $n$ qubits is $H =  E_{0} - H_{C}$, where
	\begin{equation}
		H_{C} =  \sum_{\langle v_1v_2\rangle \in E} \frac{\omega_{v_1v_2}}{2} Z_{v_1}Z_{v_2} \; .
		\label{eq:H}
	\end{equation}
	The constant
	$E_0 = \sum \omega_{v_1v_2}/2$
	plays no role in the partition of the graph, but enters the calculations of the accuracy $r$ as defined above. The ground state has an obvious $\mathbb{Z}_2$ symmetry. The ground state of $H_C$ in Eq.~\eqref{eq:H} encodes the solutions to the original MaxCut problems. We consider weighted 3-regular graphs with $\omega_{v_1v_2}$ chosen uniformly at random in $[0,1]$ (w3r graphs), and unweighted 3-regular graphs with $\omega_{v_1v_2}=1$ (u3r graphs). An example of an unweighted graph is shown in Fig.~\ref{fig:maxcut}.

	The choice of problems is the same as in Ref.~\cite{lukin}. Classically, finding a solution where $r > 16/17 \approx 0.9412$ on all graphs is NP-hard~\cite{szegedy,hastad}, but there is a polynomial time classical algorithm that provably finds answers with $r=0.8785$ \cite{goemans}.  
	
	\subsection{QAOA and ab-QAOA}\label{sec:ab_qaoa}
	
	The quantum part of the standard QAOA is the repeated computation of a quantity $|\psi_{f}^\mathrm{s}\rangle$ according to  
	\begin{equation}
		\label{eq:s}
		|\psi_{f}^{\mathrm{s}} \rangle=
		\prod_{k=1}^{p}
		\mathrm{e}^{-i\beta_{k}H_{M}^\mathrm{s}} \mathrm{e}^{-i\gamma_{k}H_{C}}|\psi_{0}^{\mathrm{s}}\rangle,
	\end{equation}
	with $H_M^\mathrm{s} = \sum_j X_j$, where $X_j$ is the Pauli $X$ matrix that acts on the $j$th qubit. $|\psi_{0}^{\mathrm{s}}\rangle$ is the ground state of $H_M^\mathrm{s}$. The operators with subscript $k$ are on the left of those with $k-1$ in $\prod_{k=1}^{p}\cdots$. The classical part is the iterative optimization of $\{\gamma_k\}$ and $\{\beta_k\}$.
	
	The ab-QAOA algorithm modifies the QAOA algorithm in the following ways,
	\begin{equation}
		\label{eq:ab}
		|\psi_{f}^{\mathrm{ab}} \rangle=
		\prod_{k=1}^{p}
		\mathrm{e}^{-i\beta_{k}H_{M}^\mathrm{ab}(\{h_j\})} \mathrm{e}^{-i\gamma_{k}H_{C}}|\psi_{0}^{\mathrm{ab}}(\{h_j\})\rangle,
	\end{equation}
	where the mixing Hamiltonian is $ H_M^\mathrm{ab} = \sum_j (X_j-h_j Z_j) $ and the starting wavefunction $|\psi_0^\mathrm{ab}\rangle$ is the ground state of the former. There are $n$ additional parameters $ \{h_j\}$ that comprise the local fields and enter both the $ H_M^\mathrm{ab}$ and  $| \psi_0^\mathrm{ab} \rangle$. They are \textit{not} optimized, but rather updated according to the prescription 
	\begin{equation}
		h_j \rightarrow h_j - \ell (h_j-\langle  \psi_f^\mathrm{ab} | Z_j  | \psi_f^\mathrm{ab} \rangle ).  
	\end{equation}
	$\ell$ is the learning rate, which we took to be $\ell = 1.1$ and $\langle  \psi_f^\mathrm{ab} | Z_j  | \psi_f^\mathrm{ab} \rangle$ can be obtained from the measurement of $ZZ$ terms in $H_C$. This is one step of the learning process. Thus both $H_M^\mathrm{ab}$ and $| \psi_0^\mathrm{ab} \rangle$ are updated (learned) along with the usual QAOA schedule parameters $\{\gamma_k\}$ and $\{\beta_k\}$ (which are optimized in the usual way at each iteration). The details of ab-QAOA in level $p$ can be found in the following.
	
	\begin{table}[H]
	    \normalsize
		\centering
		\begin{tabular}{p{\linewidth}}
			\toprule
			\textbf{Algorithm: ab-QAOA in level $p$} \\
			\hline
		\end{tabular}
	\begin{itemize}
		\item Initialization
		\begin{enumerate}
			\item Initialize 2 $p$-element sets $\{u_l\}$ and $\{v_l\}$ that are used to update $\{\gamma_k\}$ and $\{\beta_k\}$.		
			\item Initialize the $n$-element local field set $\{h_j\}$.
			\item Set a learning rate $\ell$, a global parameter defined in Step 6 in optimization procedure.
		\end{enumerate}
	\end{itemize}
	
	\begin{itemize}
		\item Optimization
		
		\begin{enumerate}
			\item Set $\{\gamma_k\}$ and $\{\beta_k\}$ according to the discrete Fourier transforms of $\{u_l\}$ and $\{v_l\}$.		
			\item Construct the mixing Hamiltonian with bias fields:
			\begin{align}
				H_{M}^\mathrm{ab}(\{h_j\}) = \sum_{j=1}^n (X_j-h_{j}Z_{j}).\nonumber
			\end{align}
			\item Prepare $|\psi_{0}^{\mathrm{ab}}\rangle$, the product ground state of $H_{M}^\mathrm{ab}(\{h_j\})$.
			
			\item Compute the final state for this step:
			\begin{equation}
			    \quad
			    \quad
				|\psi_{f}^{\mathrm{ab}} \rangle=
				\prod_{k=1}^{p}
				\mathrm{e}^{-i\beta_{k}H_{M}^\mathrm{ab}(\{h_j\})}
				\mathrm{e}^{-i\gamma_{k}H_{C}}|\psi_{0}^{\mathrm{ab}}\rangle\nonumber.
			\end{equation}
			
			\item Using projective measurements, obtain the gradients of the energy:
			\begin{equation}
				\quad
				\quad
				\frac{\partial \langle\psi_{f}^{\mathrm{ab}}| H_{C}|\psi_{f}^{\mathrm{ab}} \rangle}{\partial \vec{u}}  \quad \textrm{and} \quad  
				\frac{\partial \langle\psi_{f}^{\mathrm{ab}}| H_{C}|\psi_{f}^{\mathrm{ab}} \rangle}{\partial \vec{v}}\nonumber
			\end{equation}
			and the quantity
			\begin{equation}
			\quad
			\quad
				\delta h_{j} = h_{j} -  \langle\psi_{f}^{\mathrm{ab}}| Z_{j}|\psi_{f}^{\mathrm{ab}} \rangle.\nonumber
			\end{equation}
			
			\item Update $\{v_l\}$, $\{u_l\}$ using the Adam gradient-based stochastic optimization algorithm~\cite{adam}.  Update $\{h_j\}$ with learning rate $\ell$ according to
			$ h_j \rightarrow h_j - \ell\delta h_j$. The update of $\{h_j\}$ feeds back into both the mixing Hamiltonian in Step 2 and the wavefunction in Step 3.  
			
			\item  Measure the expectation value of the energy/cost function 
			$E(\{u_l\}, \{v_l\}, \{h_j\}) = \langle\psi_{f}^{\mathrm{ab}}| H_{C} |\psi_{f}^{\mathrm{ab}} \rangle $.
			
			\item Repeat steps 1-7 until convergence with a fixed tolerance. Output the final energy $E_{f}(\{u_l\}, \{v_l\}, \{h_j\})$, and a measurement of $|\psi_{f}^{\mathrm{ab}} \rangle$ in the computational basis. Allowing for the constant term, the optimized energy is $E^{\mathrm{opt}}_p=E_0-E_f$.
		\end{enumerate}	
	\end{itemize}
	    \centering
		\begin{tabular}{p{0.99\linewidth}}
		\quad\\
		\hline
		\end{tabular}
	\end{table}	
    
    \quad\\
    
    Besides the optimization of $\{\gamma_k\}$ and $\{\beta_k\}$, another important issue is the choice of the initial $\{\gamma_k\}$ and $\{\beta_k\}$ at the beginning of the optimization. For this, we adapt the Fourier strategy ~\cite{lukin}, as described in Appendix~\ref{sec:abQAOA Details}. The main idea is not to directly optimize $\{\gamma_k\}$ and $\{\beta_k\}$, but rather to optimize their Fourier components $\{ u_l\}$ and $\{v_l\}$, given by
	
	\begin{equation}
		\begin{split}
			\gamma_{k} &=\sum_{l=1}^{p} u_{l} \sin \left[\left(l-\frac{1}{2}\right)\left(k-\frac{1}{2}\right) \frac{\pi}{p}\right],\\
			\beta_{k} &=\sum_{l=1}^{p} v_{l} \cos \left[\left(l-\frac{1}{2}\right)\left(k-\frac{1}{2}\right) \frac{\pi}{p}\right].
		\end{split}\label{eq:uv}
	\end{equation}
	Then the starting point in level $p$ can be constructed from the optimized point in level $p-1$. 
	We note that since QAOA is the $\ell \rightarrow 0$ and $h_j \rightarrow 0$ limit of the ab-QAOA, performance guarantees for the QAOA \cite{farhi,crooks,wurtz} apply also to the ab-QAOA.    
	
	\section{numerical results}\label{sec:result}
	\subsection{Comparison between QAOA and ab-QAOA}\label{sec:energy_fidelity}
	Our primary figure of merit is the time taken to reach a given accuracy $r^*$. The choice of a target accuracy $r^*$ is to some extent arbitrary.  We will take $r^*=0.99$ as a value that is attainable in numerical simulations at moderate system sizes for the ab-QAOA, and for the QAOA with reasonable extrapolations.  This value of $r^*$ also sets a goal that may be practical for future quantum computers in the medium term and it exceeds the NP-hard threshold quoted above. The ratio of computation times for the QAOA and the ab-QAOA to reach the accuracy $r^*$ is then our measurement of the improvement in the algorithm.  We define $p^*$ as the value of the level at which $r^*$ is achieved. 
	
	To understand the dependence of the runtime on the level $p$, consider the optimization over a $p$-level output state from either QAOA or ab-QAOA. In the gradient-based classical optimization algorithm, $O(p)$ gradients are necessary and for the calculation of each gradient, a $p$-level output state needs to be prepared. Intuitively, the total quantum computation time is $O(p^2)$. which will be analyzed rigorously in Appendix~\ref{sec:time}. 
	
	Crucially, there is \textit{no} additional quantum overhead in the ab-QAOA since \textit{no} additional measurements are needed, and the number of gates for the state preparation \textit{is the same} as in the QAOA. There is classical overhead due to the larger number of parameters.  However, this cost turns out to be very small, owing to the fact that the only really important additional parameter is the bias field.	This field is not optimized over, but rather simply fed back at each iteration, and the total number of these fields is only $n$, the number of qubits. These cost issues are treated in more detail in Appendix~\ref{sec:time}. Given these considerations, the speedup is best defined as $S = (p^*_{\mathrm{QAOA}}/p^*_{\mathrm{ab-QAOA}})^2$.  We will also plot the infidelity $ 1- F = 1 - \sum_\alpha|\langle \psi_f^{\mathrm{s}(\mathrm{ab})} | \psi_{\max}^\alpha \rangle |^2$ to compare the two methods, (where $\alpha$ labels the degeneracy), since this quantity gives additional physical insight.
	
	The results of the comparison of the ab-QAOA and QAOA algorithms for w3r graphs with $n=8,12,16$ are shown in log-linear plots in Figs.~\ref{fig:w3ra}, \ref{fig:w3rb} for $1-r$ and $1-F$ respectively, while the results for $n=10,14,18$ are given in Appendix~\ref{sec:result_101418}. The convergence to the solution is much better in the ab-QAOA overall, in some cases by more than an order of magnitude.  The improvement at small $p$ is particularly striking.  This is important, since only rather small values of $p$ are likely to be accessible in near-term quantum machines~\cite{harrigan2021quantum,otterbach2017unsupervised,qiang2018large,pagano2020quantum,willsch2020benchmarking,abrams2019implementation,bengtsson2020improved}. 
	
	\begin{figure}[ht]
		\centering
		\subfigure
		{
			\includegraphics[scale=0.5]{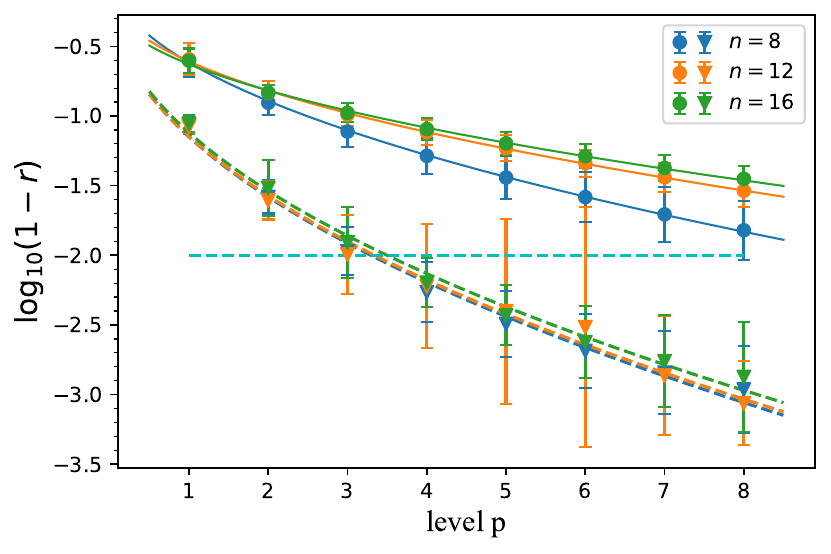}
			\label{fig:w3ra}
		}
		{
			\includegraphics[scale=0.5]{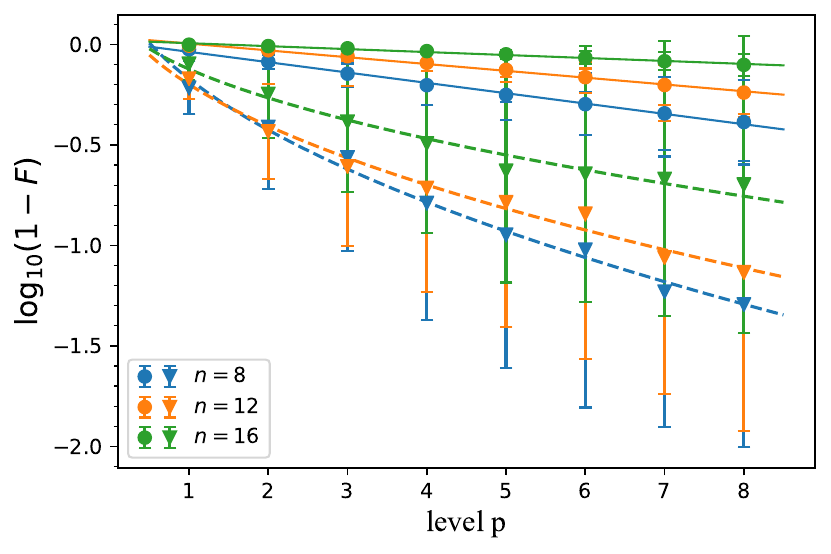}
			\label{fig:w3rb}
		}

		\caption{Comparison of the accuracy (top panel) and infidelity (bottom panel) of the QAOA (solid lines) to the accuracy and infidelity of ab-QAOA (dashed lines) for $n=8,12,16$ for w3r graphs.  Each point is an average over $40$ randomly chosen graphs.  In (a) the accuracy is plotted as a function of the level $p$.  The horizontal dashed line represents $r^*=0.99$.  Even for moderate values of $p$, the accuracy of ab-QAOA is an order of magnitude better than that of QAOA.  In (b) the infidelity in QAOA and ab-QAOA is plotted as a function of $p$.  Again, the improvement is nearly an order of magnitude at moderate $p$. The fits are described in the text. The error bars are standard deviations.}\label{fig:w3r}
	\end{figure}
	
	It is at first sight surprising that the computed accuracy is not always significantly better for smaller graphs, as seen in Fig.~\ref{fig:w3ra} for $p = 7,8$. This is due to the larger error bars at larger $p$ but the bars are magnified by the log scale. This inversion is discussed in more detail in Appendix~\ref{sec:result_101418}, where more extensive calculations are also presented.

	In order to calculate the speedup we need $p^{*}$. However, in the QAOA for larger $n$ values, the algorithm does not achieve the desired accuracy $r^{*}$ for $p \leq 8$.  Thus some extrapolation is required and this means choosing some fitting functions for $r(p)$, choosing the point where the curve intersects $r^{*}$ and rounding $p$ at that point to the nearest integer. We fit both the w3r results and the u3r results in the standard QAOA using the purely empirical forms in~\cite{lukin}.

	The fits are surprisingly good.  We have no good explanation for this at this point, but high-quality empirical fits often lead to later insights. We have also performed a scaling analysis, given in Appendix~\ref{sec:fit}, which shows that the points collapse onto a straight line in a rescaled plot. For the QAOA and w3r graphs the fitting functions are:
	\begin{equation}
		\begin{split}
			1-r &=\mathrm{exp}(-\sqrt{p/p_0}+c),\\
			1-F &=\mathrm{exp}(-p/p_0+c).
		\end{split}
	\end{equation}
	The forms for the ab-QAOA for w3r graphs are slightly different, though we do not know at this point if the difference in the forms has any fundamental significance.  The functions are:
	\begin{equation}\label{equ:ab_fit}
		\begin{split}
			1-r &=\mathrm{exp}(-\sqrt{p/p_0}+c),\\
			1-F &=\mathrm{exp}(-\sqrt{p/p_0}+c).
		\end{split}
	\end{equation}
	The fitting parameters $p_0$, $c$ and the fitting errors are tabulated in Appendix~\ref{sec:fit}.
	
	\begin{figure}[h]
		\centering
		\includegraphics[scale=0.5]{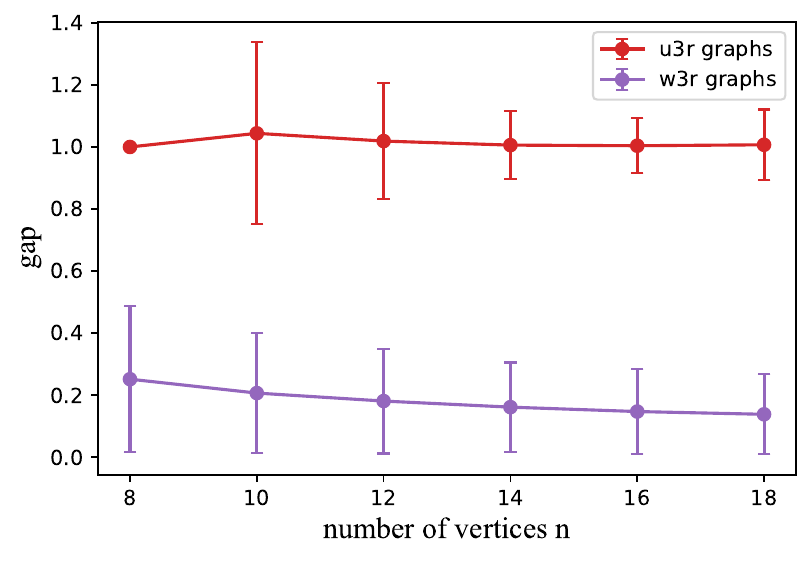}
		
		\caption{The gap between the ground state and the first excited state of unweighted or weighted graphs for $n=8,10,12,14,16,18$. Each point is the average over $1000$ graphs except $n=8$ u$3$r graphs, where there are only $5$ different non-isomorphic graphs. The error bars are standard deviations. }\label{fig:gap}
	\end{figure}

	For the w3r results, the fitting functions work very well, as can be seen in Fig.~\ref{fig:w3r}.  The upward curvature in the ab-QAOA fits is due to the fact that at higher $p$ we are close to converging to the actual solution.  It is notable that for the ab-QAOA the curvature does not increase very rapidly with $p$, indicating that even when the ab-QAOA is quite close to the actual result, improvement still continues. The results for the relative infidelity of the QAOA and the ab-QAOA are nearly as impressive as those for the accuracy; the gap between the two methods is still clearly evident. In the ab-QAOA, $1-r$ is nearly independent of $n$, while $1-F$ changes noticeably.  This is an indication that the energy spectrum of  weighted graphs differs from that of unweighted graphs: the ground state for weighted graphs is more likely to be nearly degenerate with the low-lying excited states.  This is shown by numerical calculation of the gap between the ground state and the first excited state as shown in Fig.~\ref{fig:gap} for both types of graph.
	
	\begin{figure}[ht]
		\centering
		\subfigure
		{
			
			\includegraphics[scale=0.5]{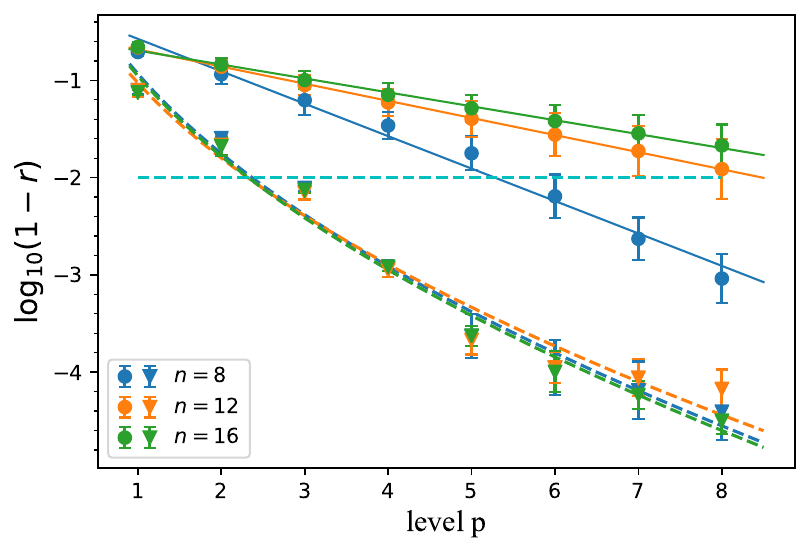}
			\label{fig:u3ra}
		}
		\subfigure
		{
			\includegraphics[scale=0.5]{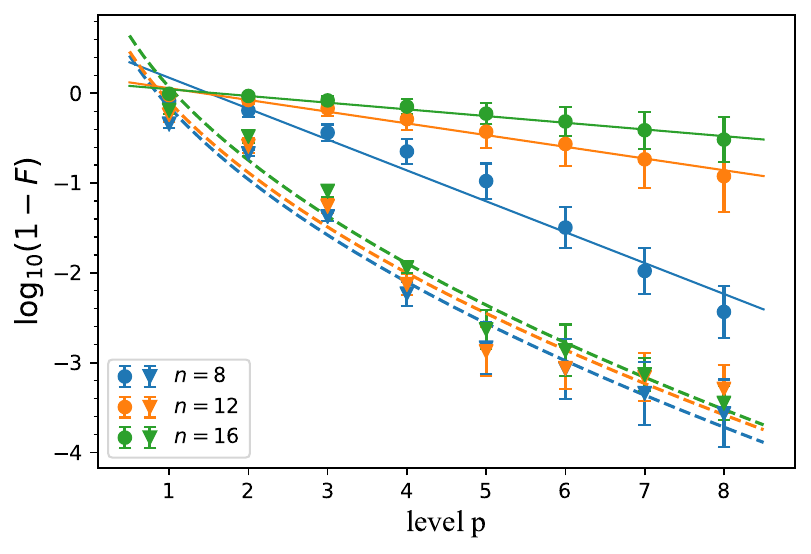}
			\label{fig:u3rb}
		}
		\caption{ Comparison of the accuracy (top panel) and the infidelity (bottom panel) for QAOA (solid lines) and ab-QAOA (dashed lines) for $n=8,12,16$ for u3r graphs.  The number of realizations is $40$ for graphs with $10$ or more vertices. For $n=8$, there are only $5$ different non-isomorphic u3r graphs, and we average over them. The error bars are standard deviations.  In (a) the accuracies for QAOA and ab-QAOA are plotted vs. $p$.  The horizontal dashed line again represents $r^*=0.99$.  By comparing with Fig.~\ref{fig:w3r} we find that the accuracy for both algorithms is slightly better when applied to  unweighted graphs. The improvement of ab-QAOA over QAOA is very marked at even smaller $p$. In (b) the infidelities for QAOA and ab-QAOA are plottted as a function of $p$, now for unweighted graphs.  Again, the improvement is clear at quite small $p$ and it continues to improve for all $p$.  The fits are described in the text. }\label{fig:u3r}
	\end{figure}
	
	The results for the u3r graphs with $n=8,12,16$ are shown in Fig.~\ref{fig:u3r} and the results for $n=10,14,18$ are given in Appendix~\ref{sec:result_101418}.  Again, the gap between the QAOA and the ab-QAOA is clearly evident.  The initial convergence at small $p$ is very fast for the ab-QAOA.  Indeed, if the figure of merit for the algorithms is taken as the accuracy at some fixed small $p$, the difference in performance for u3r graphs would exceed that for w3r graphs.

	For u3r graphs the fitting functions for QAOA are straight lines on the log-linear plots in Figs.~\ref{fig:u3ra}, \ref{fig:u3rb}:
	\begin{align}
		\begin{split}
			1-r &=\mathrm{exp}(-p/p_0+c),\\
			1-F &=\mathrm{exp}(-p/p_0+c),
		\end{split}
	\end{align}
	while for ab-QAOA, the fitting functions are the same as those in w3r graphs, those in Eq.~\eqref{equ:ab_fit}.  Again, $p_0$ and $c$ are fitting parameters that are given in Appendix~\ref{sec:fit} with fitting errors.  The fits are generally good with two slight exceptions.  The first is when  $n=8$, for which there are few graphs so that little averaging can be performed.  The second is the  ab-QAOA at large $p$, where there is additional curvature that is not captured by the fits.  In this region the accuracy is so high that the curves must flatten out and this introduces some finite-size error. Interestingly, the convergence rate of the infidelity of the ab-QAOA for u3r wavefunctions is considerably faster than that for the w3r case, and $1-F$ does not depend so strongly on $n$.  This is the main difference in our results for the w3r and u3r graphs.  We believe that this is due to the fact that for the u3r graphs the ground state is well-separated in energy from the low-lying excited states relative to the w3r graphs.  This follows from the fact that the weight parameters in $H_C$ for the u3r graphs are integers but those in the w3r graphs are not.  
	
	\begin{figure}[ht]
		\centering 
		\includegraphics[scale=0.5]{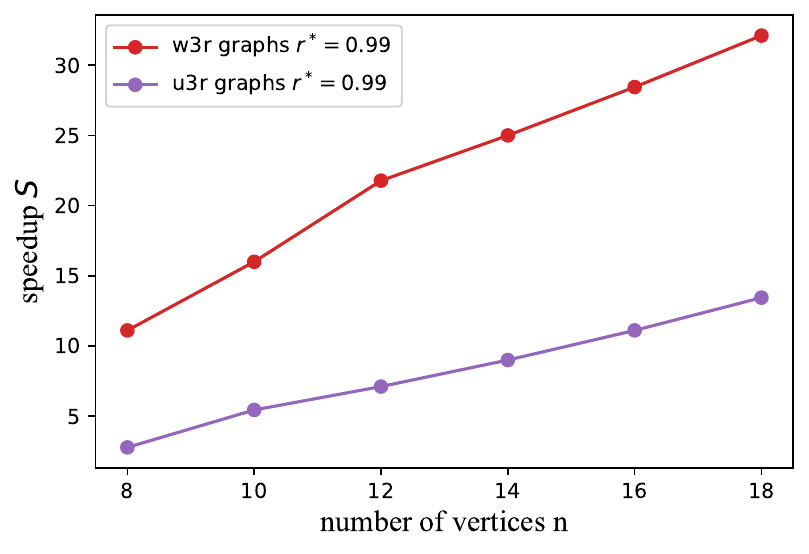} 	
		\caption{Speedup $S(n)$ of ab-QAOA over standard QAOA, using  an accuracy $r^*=0.99$ as a criterion. The corresponding $p^*$ values are given in Appendix~\ref{sec:p*}}\label{fig:speedup}
	\end{figure}

	The basic figure of merit for the ab-QAOA is $S(n)$, the speedup as a function of the number of vertices.  This is plotted in Fig.~\ref{fig:speedup}.  We see first of all that the improvement offered by the ab-QAOA is certainly not limited to very small graphs.  If $S(n)=O(n^a)$, where $a>0$, then the ab-QAOA gives a polynomial speedup over QAOA.  If $a=0$ then we can only hope for a constant speedup (which might still be of practical importance, of course).  The curve in Fig.~\ref{fig:speedup} shows no sign of saturating up to $n=18$.  The numerical results give evidence that the ab-QAOA gives a significant speed-up for the solution of the MaxCut problem compared to other classical-quantum hybrid optimization algorithms. We expect that the ab-QAOA can also be applied to other classical combinatorial optimization algorithms that can be mapped into classical Ising models, such as partitioning problems, covering and packing problems and so on~\cite{lucas2014ising,glover2019quantum}, since then the local field idea can be suitably modified.  For problems that lack such a mapping such as finding the ground states of molecules~\cite{cerezo2020variational}, the applicability of the ab-QAOA is much less clear.
	
	\subsection{Effect of bias fields for level $p=1$.}
	
\begin{figure}[h]
		\centering
		\includegraphics[scale=0.4]{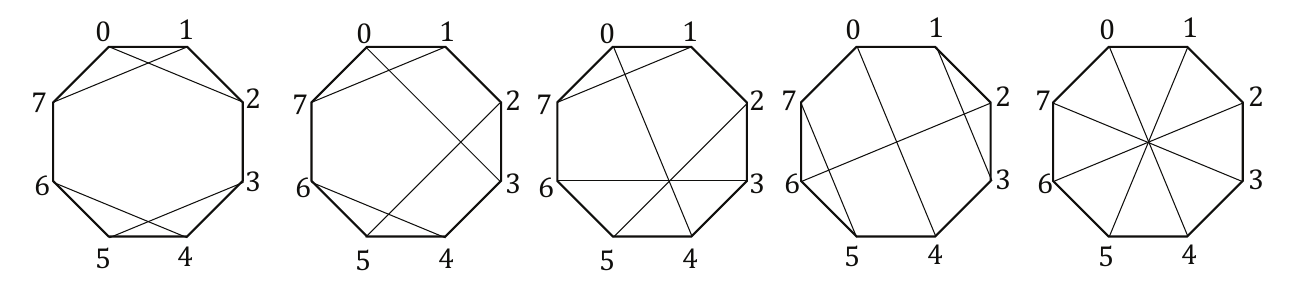}
		\caption{All the $5$ different u3r graphs with 8 vertices. These graphs are labeled as $1,2,\cdots$ in sequence.}\label{fig:u3r8}
	\end{figure}

	In this section, we illustrate the effect of bias fields for the smallest non-trivial graphs (the first graph in Fig.~\ref{fig:u3r8}) and only at level $p=1$. This isolates the effect of having these fields in the mixing Hamiltonian.  We simply repeat the evolution, measuring $\langle H_C\rangle$ at the end of each step and then update the bias fields in the mixing Hamiltonian and the starting wavefunction using the prescription given above.  Of course this leads to lower fidelities than for the full algorithm presented in Sec.~\ref{sec:energy_fidelity}.   
	
	In the ab-QAOA, the mixing Hamiltonian with bias field is:
	\begin{equation}
		H_{M}^{\mathrm{ab}}(\{h_j\}) = \sum_{j=1}^{n} (X_{j}-h_{j}Z_{j}).
	\end{equation}
	
	Once we know one product ground state $|\psi_{\max}^{\alpha}\rangle$ of the MaxCut problem Hamiltonian in Eq.\eqref{eq:H}  (whose ground states are always degenerate and $\alpha$ is used to eliminate this degeneracy), then we have expectation value of each $Z_j$. If $h_j$ is fixed to $\langle \psi_{\max}^{\alpha}|Z_j|\psi_{\max}^{\alpha}\rangle$ in our ab-QAOA, then  $|\psi_{0}^{\mathrm{ab}}\rangle$ is closer to $|\psi_{\max}^{\alpha}\rangle$ than $|-\rangle^{\otimes n}$ (the starting state of the standard QAOA), leading to a higher accuracy for the ab-QAOA.
	
	The bias field parameter $h_j$ is updated according to  
	\begin{align}
		h_j \rightarrow h_j-\ell(h_j-\langle Z_j \rangle).
	\end{align}
	This update strategy will bring $h_{j}$ closer to $\langle \psi_{\max}^{\alpha}|Z_{j}|\psi_{\max}^{\alpha}\rangle$ and the starting state $|\psi_{0}^{\mathrm{ab}}\rangle$ closer to  $|\psi_{\max}^{\alpha}\rangle$. In realistic calculations, prior knowledge of $|\psi_{\max}^{\alpha}\rangle$ may not be available. 
	It turns out that we can still find  $|\psi_{\max}^{\alpha}\rangle$ faster than the QAOA even without prior knowledge of $|\psi_{\max}^{\alpha}\rangle$, as we now show.

\begin{figure}[ht]
		\centering    
		\subfigure
		{
			\includegraphics[scale=0.48]{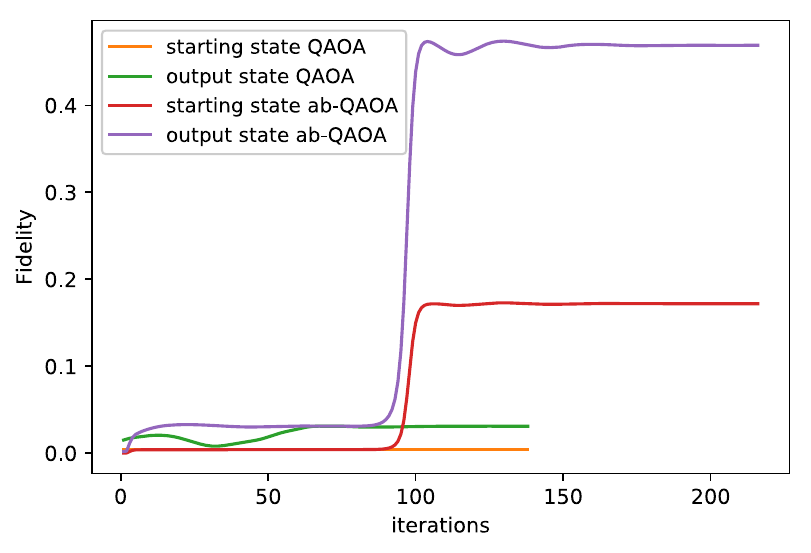}
			\label{fig:f1}
		}\\
		\subfigure
		{
			\includegraphics[scale=0.48]{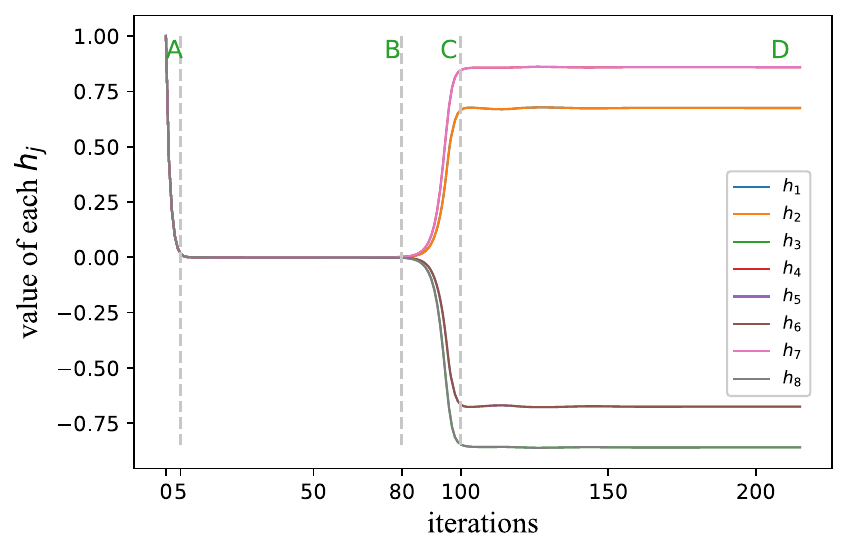} 
			\label{fig:h1}
		}

		\caption{The fidelity (top panel) between the target ground state and the starting or output states for graph 1 in Fig.~\ref{fig:u3r8} using level-$1$ QAOA and level-$1$ ab-QAOA, and the corresponding ab-QAOA bias fields (bottom panel), all plotted against the number of iterations. The target state is $|\psi_{\max}^{\alpha}\rangle$ = $|00101101\rangle$. The fidelities show sudden jumps which correspond to rapid changes in the the bias fields of the ab-QAOA. The fields steer the starting state to the desired output state.  $h_1$, $h_2$, $h_4$ and $h_7$ are greater than $0$ at the end of the optimization, corresponding to the form of $|\psi_{\max}^{\alpha}\rangle$ . The ranges A, B, C, and D are discussed in the text. }\label{fig:bias_1}
		
	\end{figure}

	To illustrate this, we calculate the fidelity $\sum_\alpha|\langle \psi_{0}^{\mathrm{ab}}|\psi_{\max}^{\alpha}\rangle|^2$, where $|\psi_{0}^{\mathrm{ab}}\rangle$ is the ground state of $H_{M}^\mathrm{ab}(\{h_j\})$ in the level-$1$ ab-QAOA for the first u3r graph with $8$ vertices as shown in Fig.~\ref{fig:u3r8}. The sum is over the ground states that ab-QAOA steers the starting state to. For comparison, we also calculate $\sum_\alpha|\langle \psi_{0}^{\mathrm{s}}|\psi_{\max}^{\alpha}\rangle|^2$,  $\sum_\alpha|\langle \psi_{f}^{\mathrm{s}}|\psi_{\max}^{\alpha}\rangle|^2$ and  $\sum_\alpha|\langle \psi_{f}^{\mathrm{ab}}|\psi_{\max}^{\alpha}\rangle|^2$ in Fig.~\ref{fig:bias_1}, where $|\psi_{0}^{\mathrm{s}}\rangle$ is the starting state of the standard QAOA,  $|\psi_{f}^{\mathrm{s}}\rangle$ is the output state produced by the QAOA and $|\psi_{f}^{\mathrm{ab}}\rangle$ is the output state produced by the ab-QAOA. 
	
	As shown in Fig.~\ref{fig:f1}, it is clear that the bias field will bring the starting state closer to the ground state of $H_{C}$. There are some iterations for which both the starting state and output state curves of ab-QAOA grow rapidly and that is when the bias field brings the starting state $|\psi_0^\mathrm{ab}\rangle$ close to the ground states.  The operations $\exp(-i\beta_k H_M^\mathrm{ab})$ and $\exp(-i\gamma_k H_c)$ bring $|\psi_f^\mathrm{ab}\rangle$ closer to the target states than $|\psi_0^\mathrm{ab}\rangle$.  Note that the fidelity approaches $0.5$ for the output state of the ab-QAOA. This is bounded above by the ab-QAOA driven by fixed bias fields $h_j=\langle\psi_{\max}^\alpha|Z_j|\psi_{\max}^\alpha\rangle$ with $\ell=0$.
		
	To better investigate how the bias fields work, we also plot $\{h_j\}$ of graph $1$ from level $1$ ab-QAOA in the  optimization iterations as shown in Fig.~\ref{fig:h1}. There are four regions in Fig.~\ref{fig:h1}.  In Region A, all $h_j$ decrease to $0$ quickly in the first 5 iterations. In Region B, from the $5$th iteration to the $80$th iteration, all  $h_j$ are near $0$. In Region C, from about the $80$th iteration to the $100$th iteration, the $\{h_j\}$ diverge and each $h_j$ tries to find its true value, $\langle\psi_{\max}^\alpha|Z_j|\psi_{\max}^\alpha\rangle$. In Region D, in the last half of the optimization, the value of each $h_j$ doesn't change. The behavior of the fidelity in Fig.~\ref{fig:f1} is related to $\{h_j\}$ in Fig.~\ref{fig:h1}. The divergence of $\{h_j\}$ from $0$ implies a sharp rise in the fidelity.

\begin{figure*}[ht]
		\subfigure
		{
			\begin{minipage}[h]{0.45\linewidth}
				\centering          
				\includegraphics[scale=0.6]{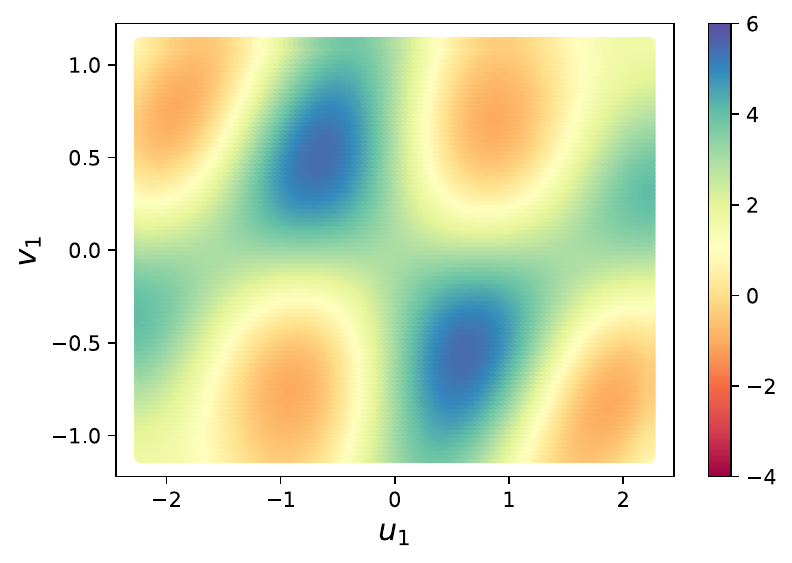}   
			\end{minipage}
		}
		{
			\begin{minipage}[h]{0.45\linewidth}
				\centering          
				\includegraphics[scale=0.6]{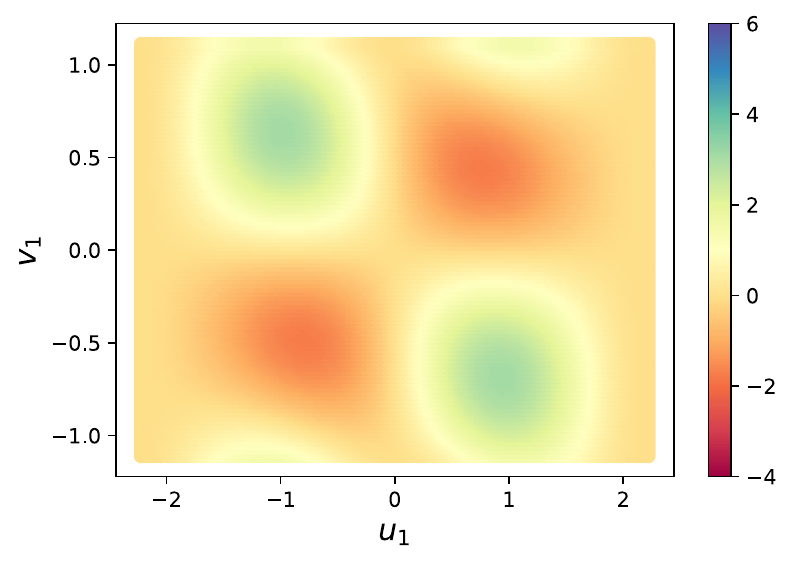}   
			\end{minipage}
		}\\
		{
			\begin{minipage}[h]{0.45\linewidth}
				\centering          
				\includegraphics[scale=0.6]{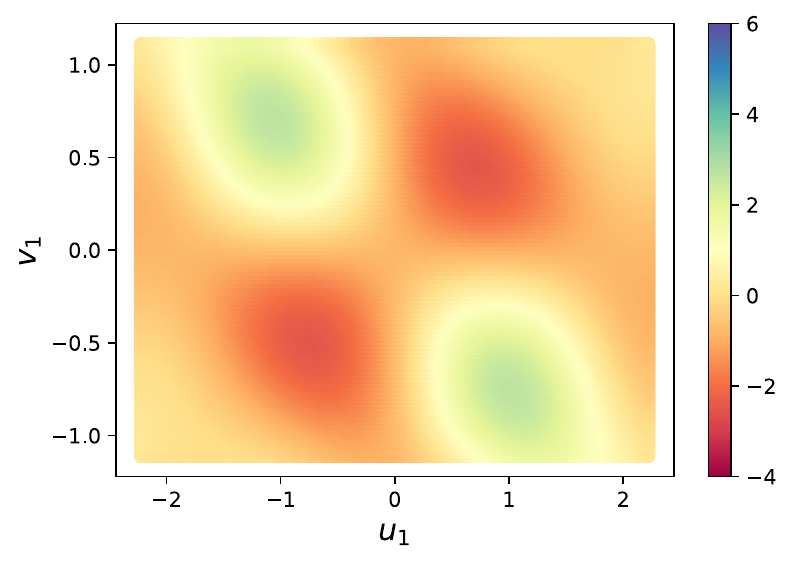}   
			\end{minipage}
		}
		{
			\begin{minipage}[h]{0.45\linewidth}
				\centering          
				\includegraphics[scale=0.6]{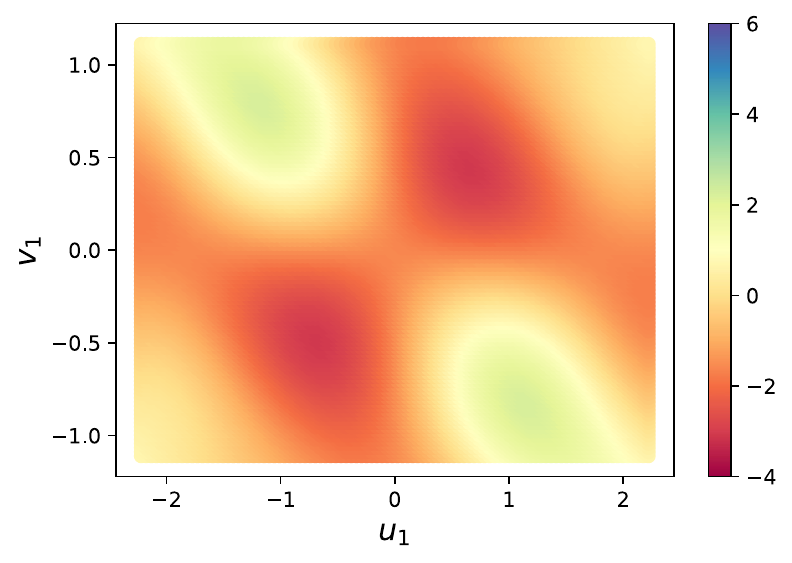}   
			\end{minipage}
		}
		
		\caption{The energy landscape of graph $1$ as a function of the variational parameters $u_1$ and $v_1$ for increasing number of iterations. The $0$th (top left), $50$th (top right), $95$th (bottom left), and final (bottom right) iterations that are shown belong to the region A, B, C and D. As analysed in Ref.~\cite{lukin}, $\gamma_1$, $\beta_1$ can be restricted to $[-\pi/2,\pi/2]$ and $[-\pi/4,\pi/4]$ respectively, so $u_1$ and $v_1$ are restricted to $[-\sqrt{2}\pi/2,\sqrt{2}\pi/2]$ and $[-\sqrt{2}\pi/4,\sqrt{2}\pi/4]$ according to Eq.\eqref{eq:uv}.  }\label{fig:landscape}
	\end{figure*}

	For each of these four regions, we choose four specific points and plot the energy landscape using ab-QAOA as shown in Fig.~\ref{fig:landscape}. Note that for region B, the landscape is in close agreement with that of the QAOA since all $h_j$ are small. In region A, due to the "wrong" bias fields, it is harder to find the target state using ab-QAOA than the QAOA, so the QAOA output state can be regarded as the optimal state of ab-QAOA's. As a result, each $h_j$ moves towards $0$ fast in region A. In region B, although all $h_j$ are small, their effects accumulate until the bias fields can have a significant effect on the cost function. In region C, all $h_j$ change quickly because of the accumulation in region B. In the updating, all $h_j$ are getting closer to their true values, so it is easier to find the target state 
	in this region, which can be verified by the smaller lowest energy in the landscape. In region D, the output energy nearly meets the convergence criterion and each $h_j$ finds its true value, $\langle\psi_{\max}^\alpha|Z_j|\psi_{\max}^\alpha\rangle$, so the lowest energy is smaller than in the other regions.

	\section{conclusion}\label{sec:conclusion}
	In this paper we have shown how a generalization of the QAOA, the ab-QAOA, can greatly reduce the depth of the quantum circuit needed to solve optimization problems to a given accuracy.  The understanding of this comes partly from a study of the effects of the bias fields on small graphs. In the short and medium term (NISQ era), the results presented in Figs.~\ref{fig:w3r} and~\ref{fig:u3r} are the most important ones.  They show that a quantum computer with of order $20$ very high-quality logical qubits may produce impressive results at level $p=5$, a machine that may be attainable quite soon ~\cite{harrigan2021quantum}. In the longer term we are more interested in how the performance of the ab-QAOA scales with $n$. Fig.~\ref{fig:speedup} shows that the speedup in fact increases as the size of the system increases, which suggests that the ab-QAOA may still be the algorithm of choice beyond the NISQ era. 

	The ability to carry out practical calculations in the NISQ era will depend on finding algorithms which can be implemented in circuits of relatively shallow 	depth, and converge quickly to an answer. The ab-QAOA contributes to these goals, converging to a desired accuracy with a computation time that is polynomially shorter in the system size than that of the standard QAOA.

	\begin{acknowledgments} 
		Yunlong Yu and Xiang-Bin Wang acknowledge National Natural Science Foundation of China Grants No. 11974204 and No. 12174215. Nic Shannon acknowledges the support of the Theory of Quantum Matter Unit, Okinawa Institute of Science and Technology Graduate University (OIST).  This research was performed using the computing resources and assistance of the UW-Madison Center For High Throughput Computing.  We thank M.G. Vavilov and B. \"{O}zguler for useful discussions.
	\end{acknowledgments}
	
	\appendix

	\section{Computational details of the ab-QAOA}\label{sec:abQAOA Details}
	In this section we give further details of the Fourier strategy proposed in~\cite{lukin} about how to choose the starting points in the optimization. For a $p$-level QAOA, as stated in the main text, the mixing Hamiltonian is $H_M^\mathrm{s}=\sum_j X_j$.	The quantum processor is initialized in $|\psi_{0}^{\mathrm{s}}\rangle$, the ground state of $H_M^{\mathrm{s}}$. Then we alternately apply problem Hamiltonian $H_C$ and mixing Hamiltonian $H_M^\mathrm{s}$ to generate the final state,
	\begin{equation}
		|\psi_{f}^{\mathrm{s}} \rangle=
		\prod_{k=1}^{p}
		\mathrm{e}^{-i\beta_{k}H_{M}^\mathrm{s}} \mathrm{e}^{-i\gamma_{k}H_{C}}|\psi_{0}^{\mathrm{s}}\rangle,
	\end{equation}
	where the level $p$ is the number of times the unitary operators corresponding to $H_M^\mathrm{s}$ and $H_C$ are applied to the initial state to move it to the final state.  The scheduling parameters $\{\gamma_k\}$, $\{\beta_k\}$ in the operators are determined by optimizing
	\begin{align}
		\langle H_C \rangle(\{\gamma_k\},\{\beta_k\})=\langle\psi_{f}^{\mathrm{s}} |H_C|\psi_{f}^{\mathrm{s}}\rangle.
	\end{align}
	
	Note that for the original QAOA~\cite{farhi,lukin}, $|\psi_0^{\mathrm{s}}\rangle$ is $|+\rangle^{\otimes n}$, but in our description, $|\psi_0^{\mathrm{s}}\rangle$ is $|-\rangle^{\otimes n}$, the ground state of $H_M^{\mathrm{s}}$. If we denote the QAOA final state from $|+\rangle^{\otimes n}$ by $|\psi_f^\mathrm{s+}\rangle$, the final state from $|-\rangle^{\otimes n}$ by $|\psi_f^\mathrm{s-}\rangle$ and define $\Tilde{Z}=\prod_j Z_j$, it is easy to prove that
	\begin{align}
		\begin{split}
			|\psi_f^\mathrm{s-}\left(\{\gamma_k\},\{\beta_k\}\right)\rangle &=
			\prod_{k=1}^{p}
			\mathrm{e}^{-i\beta_{k}H_{M}^\mathrm{s}} \mathrm{e}^{-i\gamma_{k}H_{C}}\Tilde{Z}|+\rangle^{\otimes n}\\
			&=\Tilde{Z}\prod_{k=1}^{p}
			\mathrm{e}^{i\beta_{k}H_{M}^\mathrm{s}} \mathrm{e}^{-i\gamma_{k}H_{C}}|+\rangle^{\otimes n}\\
			&=\Tilde{Z}|\psi_f^\mathrm{s+}\left(\{\gamma_k\},\{-\beta_k\}\right)\rangle.
		\end{split}
	\end{align}

	There is no difference in the classical optimization for both $|\psi_f^\mathrm{s-}\rangle$ and $|\psi_f^\mathrm{s+}\rangle$ as long as $H_C$ is a classical Ising Hamiltonian since 
	\begin{align}
	    \langle \psi_f^\mathrm{s-} |H_C|\psi_f^\mathrm{s-}\rangle (\{\gamma_k\},\{\beta_k\})=\langle \psi_f^\mathrm{s+} |H_C|\psi_f^\mathrm{s+}\rangle (\{\gamma_k\},\{-\beta_k\}).
	\end{align}
	
	There are two differences in the quantum part of the two algorithms.
	\begin{enumerate}
	 
	\item In the $p$-level ab-QAOA, $H_M^\mathrm{ab}$ contains local longitudinal fields as well as the usual global transverse field, $H_{M}^\mathrm{ab}(\{h_j\}) = \sum_{j=1}^n (X_j-h_{j}Z_{j})$.
	\item The wavefunction at the initial stage of each learning step is chosen to be the ground state of the updated $H_M^\mathrm{ab}$.
	\end{enumerate}
	Hence both the longitudinal fields in the mixing Hamiltonian  and the "re-initialized" wavefunction change during the course of the ab-QAOA algorithm.

    Thus the final state of ab-QAOA is
	\begin{equation}
		|\psi_{f}^{\mathrm{ab}} \rangle=
		\prod_{k=1}^{p}
		\mathrm{e}^{-i\beta_{k}H_{M}^\mathrm{ab}(\{h_j\})}
		\mathrm{e}^{-i\gamma_{k}H_{C}}|\psi_{0}^{\mathrm{ab}}(\{h_j\})\rangle.\label{eq:psiab}
	\end{equation}
	
	In the Fourier strategy~\cite{lukin}, what we optimize over are $\{u_l\}$ and $\{v_l\}$, the Fourier components of the Fourier components of $\{\gamma_{k}\}$ and $\{\beta_{k}\}$. To avoid being trapped in the local optimum as far as possible, we start the optimization from $R$ initial points and find the point with the best energy, as was done in Ref. \cite{lukin}. To reach level $p$, we start from level $1$ and find the point with the best energy from $R$ initial points after the optimization. In level-2, $R$ initial points are generated by adding some random numbers to the best point in level $1$. Then repeat the optimization and initial point generation procedure with increasing level $p^\prime$ until $p^\prime=p$.

	Here, we use $p$ to represent the target level and $p^\prime$ to represent the inner levels. The whole process from level $1$ to level $p$ including the initial points generation is the outer loop of the algorithm while the update of $\{u_l\},\{v_l\},\{h_j\}$ until convergence for a fixed level $p^\prime$ (the algorithm presented in Sec.\ref{sec:ab_qaoa}) is the inner loop of the algorithm. The same loops are used for the QAOA except that all the $h_j$ are set to $0$. 

\begin{figure*}[ht]
		\centering    
		\includegraphics[scale=0.45]{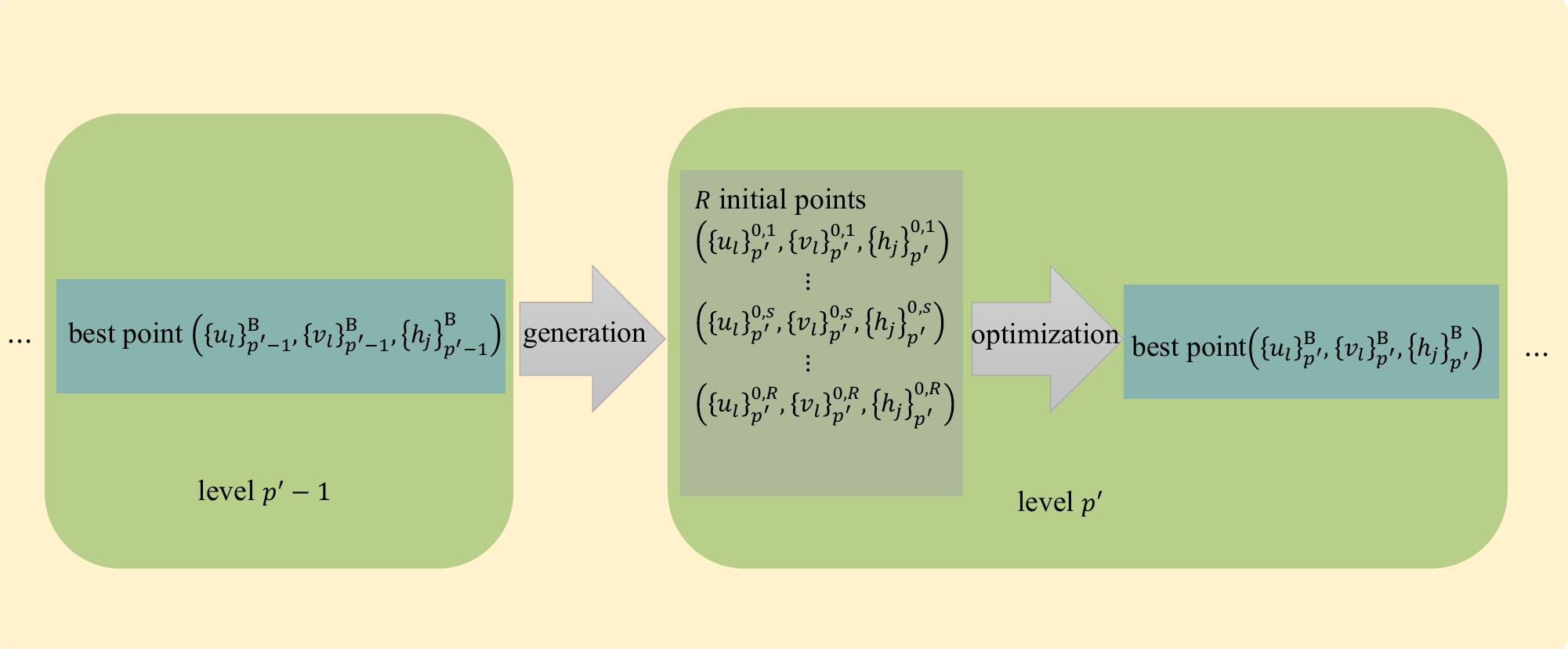} 
		\caption{Schematics of the outer loop of ab-QAOA. Using Eq.~\eqref{update}, we generate $R$ initial points in level $p^\prime$ from the best point in level $p^\prime-1$. After the optimization of these $R$ points, we get the point with the best energy. We do this procedure iteratively until the target level $p$.}\label{fig:update}  
	\end{figure*}

	This more elaborate classical optimization is not strictly necessary to demonstrate the advantages of the ab-QAOA over the QAOA, but it does mean that the results can be compared more directly with those of  Ref. \cite{lukin}. The sampling parameter $R$ was set to $10$ in our calculations. We note once more that formally the QAOA can be considered as the limit of the ab-QAOA when $h_j \rightarrow 0$. This means that from a formal standpoint the ab-QAOA is guaranteed to be at least as good as the QAOA. The detailed outer loop of the algorithm follows and is also illustrated in Fig.~\ref{fig:update}.

	\begin{table}[ht]
    	\normalsize
		\centering
		\begin{tabular}{p{\linewidth}}
			\toprule
			\textbf{Algorithm: Outer Loop from level $1$ to $p$} \\
			\hline
		\end{tabular}
		
	\begin{enumerate}
		\item In level $1$, we generate $R$ initial "$0$" points $\left(\{u_l\}_1^{0,s},\{v_l\}_1^{0,s},\{h_j\}_1^{0,s}\right)$, where the elements of $\{u_l\}_1^{0,s}$ and $\{v_l\}_1^{0,s}$ are random numbers drawn from a uniform distribution and all elements of $\{h_j\}_1^{0,s}$ are initialized to be $1$.  The subscripts refer to the ab-QAOA level in the outer loop, and the $s$ superscript ranges from $1$ to $R$ representing the different initial points. Using the algorithm in Sec.~\ref{sec:ab_qaoa} we get the optimal "$\mathrm{B}$" point $\left(\{u_l\}_1^{\mathrm{B}},\{v_l\}_1^{\mathrm{B}},\{h_j\}_1^{\mathrm{B}}\right)$ with the best optimal energy $E_1^\mathrm{B}$ from $R$ points for this level. 
		\item \label{alg}In level $p^\prime$ greater than 1, we use the best point $\left(\{u_l\}_{p^\prime-1}^{\mathrm{B}},\{v_l\}_{p^\prime-1}^{\mathrm{B}},\{h_j\}_{p^\prime-1}^{\mathrm{B}}\right)$ in level $p^\prime-1$ to construct $R$ initial points $\left(\{u_l\}_{p^\prime}^{0,s},\{v_l\}_{p^\prime}^{0,s},\{h_j\}_{p^\prime}^{0,s}\right)$. The $s$ superscript refers to the elements of the following random selection procedure, representing the different points.
		\begin{equation}\label{update}
			\begin{split}
				\{u_l\}_{p^\prime}^{0,s}&=\left\{
				\begin{array}{ll}
					\{u_l\}_{p^\prime-1}^{\mathrm{B}}\cup\{0\},&s=1\\
					\{u_l+\alpha \mathrm{Ran}^s[u_l]\}_{p^\prime-1}^\mathrm{B}\cup\{0\},&2\leq s\leq R
				\end{array}
				\right.\\
				\{v_l\}_{p^\prime}^{0,s}&=\left\{
				\begin{array}{ll}
					\{v_l\}_{p^\prime-1}^{\mathrm{B}}\cup\{0\},&s=1\\
					\{v_l+\alpha \mathrm{Ran}^s[v_l]\}_{p^\prime-1}^\mathrm{B}\cup\{0\},&2\leq s\leq R
				\end{array}
				\right.\\
				\{h_j\}_{p^\prime}^{0,s}&=\left\{
				\begin{array}{ll}
					\{h_j\}_{p^\prime-1}^{\mathrm{B}},&s=1\\
					\{h_j+\alpha \mathrm{Ran}^s[h_j]\}_{p^\prime-1}^\mathrm{B},&2\leq s\leq R
				\end{array}
				\right.\\
			\end{split}
		\end{equation}
		$\{u_l\}_{p^\prime}^{0,s}$ or $\{v_l\}_{p^\prime}^{0,s}$ is a $p^\prime$-element set whose $p^\prime$th element is zero. The random number $\mathrm{Ran}^s[a]$ is the $s$-th selection from a normal distribution with average $0$ and variance $a^2$, \textit{i.e.}, $\mathrm{Ran}^s[a]=\mathrm{Norm}(0,a^2)$. We optimize these $R$ initial points to find the best point $\left(\{u_l\}_{p^\prime}^{\mathrm{B}},\{v_l\}_{p^\prime}^{\mathrm{B}},\{h_j\}_{p^\prime}^{\mathrm{B}}\right)$ with the best energy $E_{p^\prime}^\mathrm{B}$.  The update parameter $\alpha$ was set to $ \alpha = 0.6$.
		\item Repeat step 2 until $p^\prime$ reaches the target level $p$.
		\item Output all energies $E_{p^\prime}^\mathrm{B}$ from level $1$ to $p$.
	\end{enumerate}	
		\begin{tabular}{p{\linewidth}}
		\quad\\
		\hline
		\end{tabular}
	\end{table}	
	
	\section{Computation time}\label{sec:time}
	Here we give the analysis that leads to the conclusion in the main text that the total computation time is $O(p^2)$.  We assume that the quantum part of the algorithm dominates the time.  This will be true for the foreseeable future.  The MaxCut cost Hamiltonian $H_{C}$ is defined on an $n$-vertex $\mathcal{R}$-regular graph, and a $p$-level QAOA and ab-QAOA are implemented with optimization to find a target state.  In our calculations in the main text $\mathcal{R} = 3$.
	\begin{figure*}[ht] 
		\centering    
		\includegraphics[scale=0.9]{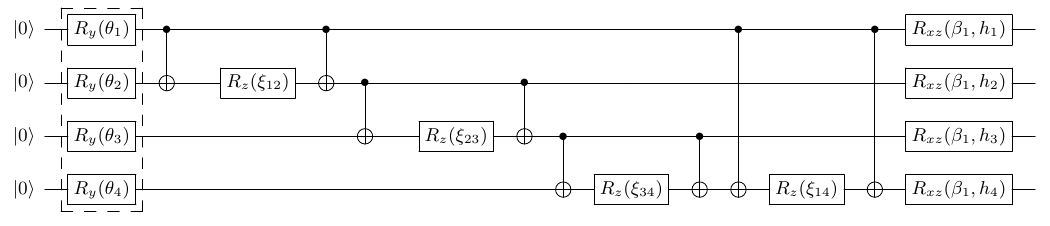} 
		\caption{Quantum circuit for $1$-level ab-QAOA on 2-regular graphs with 4 vertices. $\xi_{v_{1}v_{2}}$ is the real coefficient of $Z_{v_{1}}Z_{v_{2}}$ appearing in $\exp(-i\gamma_1 H_C)$. $R_y$ and $R_z$ are the rotation operators around the $\hat{y}$ and $\hat{z}$ axis respectively while $R_{xz}(\beta_1,h_j)=\exp[-i\beta_1 (X_j-h_j Z_j)]$. When $h_j=0$, $R_{xz}$ is the rotation operator around the $\hat{x}$ axis. The gates in the dashed box prepare the starting state for ab-QAOA. There are $4+3\times 4+4=20$ gates in the circuit. }\label{fig:circuit}
	\end{figure*}
	We denote the iterations needed for convergence by $N_{\mathrm{ite}}$. In each iteration of the optimization in our calculation, we need to calculate the expectation of the problem Hamiltonian $\langle H_C\rangle$ $2p+1$ times to get gradients of the input parameters. In both of these two QAOA, the gradient of $E_p$ (the energy in one iteration for the $p$ level QAOA or ab-QAOA) with respect to the $u_{l^\prime}$ is  
	\begin{equation}
		\begin{split}
			\frac{\partial E_p}{\partial u_{l^\prime}}&=\frac{E_p(\{u_l\}^\prime,\{v_l\},\{h_j\})-E_p(\{u_l\},\{v_l\},\{h_j\})}{\epsilon_g},\\
			\{u_l\}^\prime&=\{u_1,u_2,\cdots,u_{l^\prime}+\epsilon_g,\cdots\},
		\end{split}
	\end{equation}
	where $\epsilon_g$ is a small quantity. There are $p$ $u_l$, so $p$ $E_p(\{u_l\}^\prime,\{v_l\},\{h_j\})$ and one $E_p(\{u_l\},\{v_l\},\{h_j\})$ are needed. As a result, $2p+1$ calculations of $\langle H_C\rangle$ are needed.

	In a single calculation of $\langle H_C\rangle$, one needs to measure $n\mathcal{R} /2$ different $ZZ$ terms of $H_C$.  $|\psi_f\rangle$ (either $|\psi_f^{\mathrm{s}}\rangle$ or $|\psi_f^{\mathrm{ab}}\rangle$ in the main text) 
	is prepared $M_{ZZ}$ times to get an accurate expectation value for the $ZZ$ term. 
	
	In the ab-QAOA, unlike the QAOA, knowledge of the $Z$ term is also needed to guide $\{h_j\}$ in the flowing iteration. However, this does not require an additional measurement, since if we have the value of $\langle ZZ\rangle$ measured in the computational basis, we automatically also know $\langle Z\rangle$, as we now show. Consider a single $ZZ$ term, $Z_{v_1}Z_{v_2}$.  It has a spectral decomposition
	\begin{equation}
		\begin{split}
			Z_{v_1}Z_{v_2}&=|0_{v_1}\rangle\langle 0_{v_1}|\otimes|0_{v_2}\rangle\langle 0_{v_2}|-|0_{v_1}\rangle\langle 0_{v_1}|\otimes|1_{v_2}\rangle\langle 1_{v_2}|\\
			&-|1_{v_1}\rangle\langle 1_{v_1}|\otimes|0_{v_2}\rangle\langle 0_{v_2}|+|1_{v_1}\rangle\langle 1_{v_1}|\otimes|1_{v_2}\rangle\langle 1_{v_2}|,
		\end{split}
	\end{equation} 
	where $|1_{v_1}\rangle\langle 1_{v_1}|\otimes|1_{v_2}\rangle\langle 1_{v_2}|$ is short for 
	$\mathbb{I}\otimes\cdots\otimes\underbrace{|1\rangle\langle 1|}_{v_1}\otimes\cdots\otimes\underbrace{|1\rangle\langle 1|}_{v_2}\otimes\cdots\otimes\mathbb{I}$, which is denoted as $T_{11}^{v_1v_2}$, so as for $T_{10}^{v_1v_2}$, $T_{01}^{v_1v_2}$ and  $T_{00}^{v_1v_2}$. Once these four $T$ operators are measured then $\langle Z\rangle$ can be obtained:
	\begin{align}
		\begin{split}
			\langle Z_{v_1}\rangle&=\langle T_{00}^{v_1v_2}\rangle+\langle T_{01}^{v_1v_2}\rangle-\langle T_{10}^{v_1v_2}\rangle-\langle T_{11}^{v_1v_2}\rangle,\\
			\langle Z_{v_2}\rangle&=\langle T_{00}^{v_1v_2}\rangle+\langle T_{10}^{v_1v_2}\rangle-\langle T_{01}^{v_1v_2}\rangle-\langle T_{11}^{v_1v_2}\rangle.
		\end{split}
	\end{align}
	As a result, there are no additional measurements needed in the ab-QAOA compared to the QAOA.
	
	In one preparation of $|\psi_f\rangle$, the operator $\exp(-i\gamma_k H_C)$ is applied $p$ times and $\exp(-i\beta_k H_M^\mathrm{s})$ or $\exp(-i\beta_k H_M^\mathrm{ab})$ is applied $p$ times.  The operator $\exp(-i\gamma_k H_C)$ can be decomposed into $3$ quantum gates while $\exp(-i\beta_k H_M^\mathrm{ab})$ can be represented by one, as shown in Fig.~\ref{fig:circuit}, so $p(3n\mathcal{R}/2+n)$ quantum gates are needed.  In the meanwhile, for ab-QAOA $n$ $R_y$ rotation gates around $\hat{y}$ axis are needed for the starting state preparation from $|0\rangle^{\otimes n}$, and for the QAOA, $n$ Hadamard gates are needed.
	
	\begin{figure*}[ht]
		\centering   
		{
			\begin{minipage}[h]{0.45\linewidth}
				\centering         
				\includegraphics[scale=0.5]{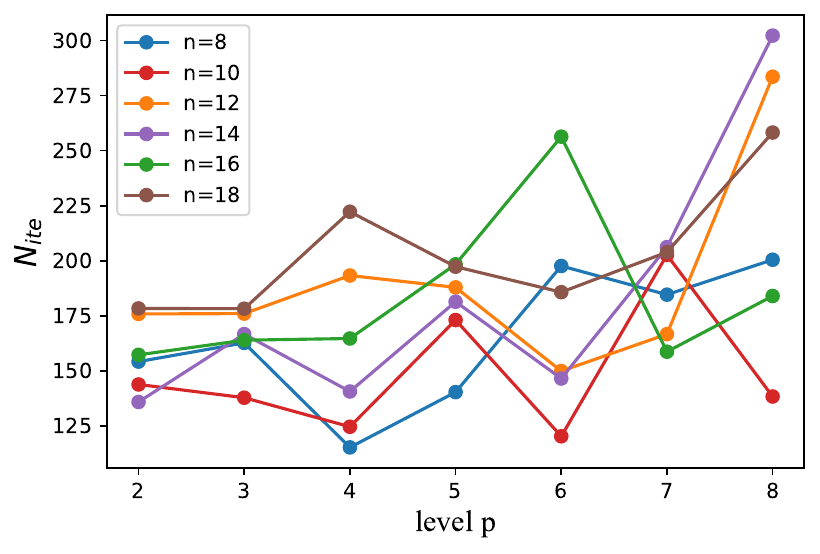}   
			\end{minipage}
		}
		{	
			\begin{minipage}[h]{0.45\linewidth}
				\centering         
				\includegraphics[scale=0.5]{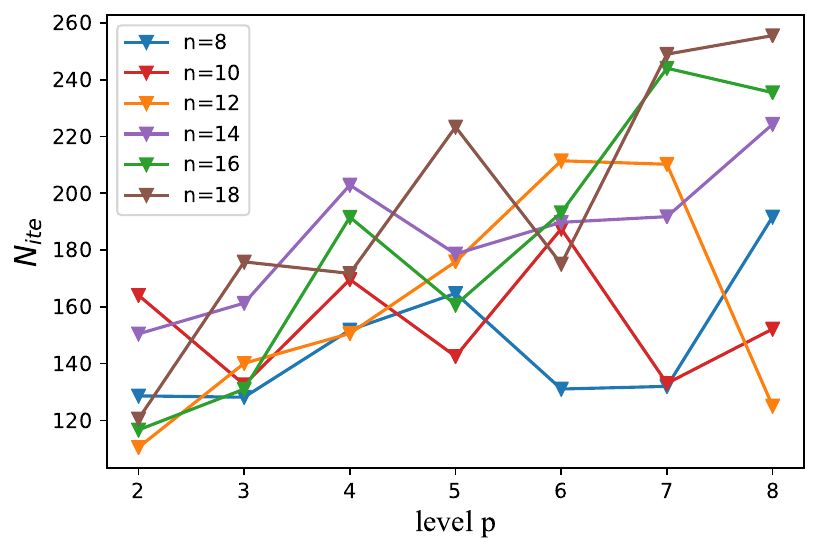}   
			\end{minipage}
		}\\
		
		{
			\begin{minipage}[h]{0.45\linewidth}
				\centering         
				\includegraphics[scale=0.5]{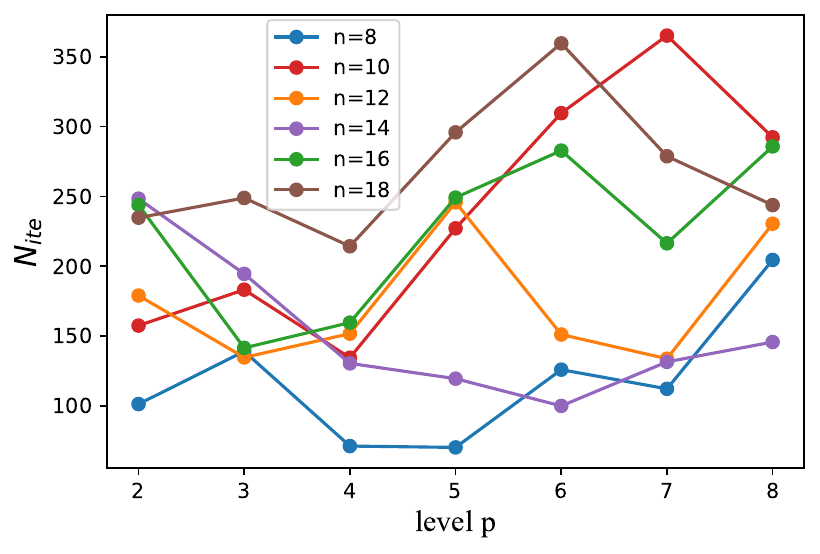}   
			\end{minipage}
		}
		{	
			\begin{minipage}[h]{0.45\linewidth}
				\centering         
				\includegraphics[scale=0.5]{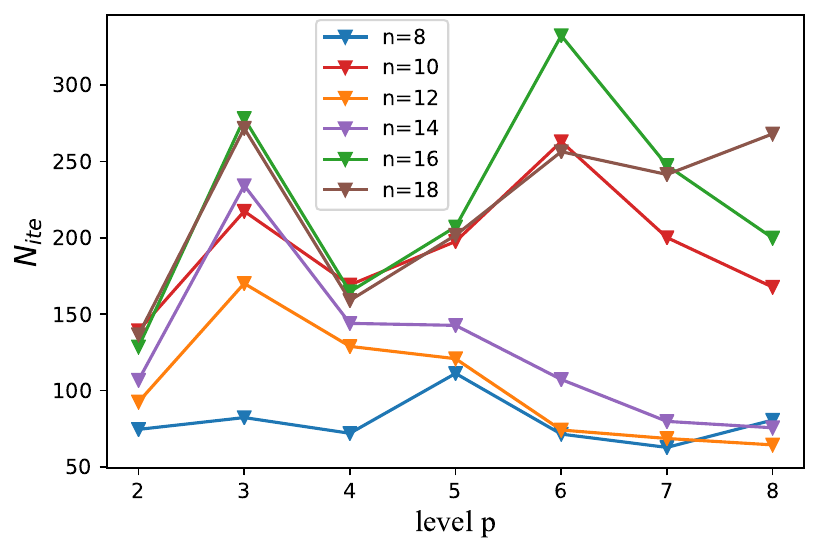}   
			\end{minipage}
		}
		
		\caption{Iterations needed for convergence $N_{\mathrm{ite}}$ in u3r graphs and w3r graphs for points generated by the above outer loop.  Top left panel is QAOA for w3r graphs, top right panel is ab-QAOA for w3r graphs, bottom left panel is QAOA for u3r graphs, and bottom right panel is ab-QAOA for w3r graphs. $N_{\mathrm{ite}}$ is the average over $R$ samples. The classical optimizer is the Adam gradient-based stochastic optimization algorithm mentioned above. $N_{\mathrm{ite}}$ is very similar for different graphs and for the two different algorithms.}\label{fig:ite_level2}
		
	\end{figure*} 
	
	In conclusion, there are
	\begin{equation}
		N_{\mathrm{gate}}=N_{\mathrm{ite}}(2p+1)M_{ZZ}\frac{n\mathcal{R}}{2}[p(\frac{3n\mathcal{R}}{2}+n)+n]
	\end{equation}
	quantum gates for a $p$-level QAOA or ab-QAOA with full optimization, $N_{\mathrm{gate}}\sim O(N_{\mathrm{ite}} p^2n^2\mathcal{R}^2)$.

	In our simulation, there are two kinds of initial points. One kind is the randomly generated points in level $1$, and the other one is the points generated with the above outer loop in the other levels. Since $p^{*}$ is always larger than $1$ for $r^{*}=0.99$, we focus on $N_{\mathrm{ite}}$ when the level $p\geq 2$. In this case, the iterations are similar among different graphs and between the QAOA and ab-QAOA as shown in Fig.~\ref{fig:ite_level2}. So we conclude $N_{\mathrm{ite}}$ is the same constant for different levels and for both algorithms, so $N_{\mathrm{gate}}\sim O(p^2n^2\mathcal{R}^2)$.  The additional classical cost for the ab-QAOA is only a small constant.  This is because essentially the entire classical cost is in the optimization routine, which does not depend on whether bias fields are included, since these fields are not optimized over.  
	
	Of course this analysis assumes that there is no error correction. It also assumes that 2-qubit gates can be applied to any pair of qubits, thus avoiding the necessity of SWAP gates.  These considerations apply equally to QAOA and ab-QAOA, so they should not affect the speedup that is defined in the main text since it is a \textit{relative} speedup.  Similarly, $\mathcal{R}$ and $n$ are the same for the two algorithms and the same reasoning may be applied.  For a given accuracy and problem size, only $p$ is different.  
	
	\section{Numerical results for $n=10,14,18$}\label{sec:result_101418}

    \begin{figure*}[!htb]
		{
			\begin{minipage}[h]{0.45\linewidth}
				\centering          
				\includegraphics[scale=0.52]{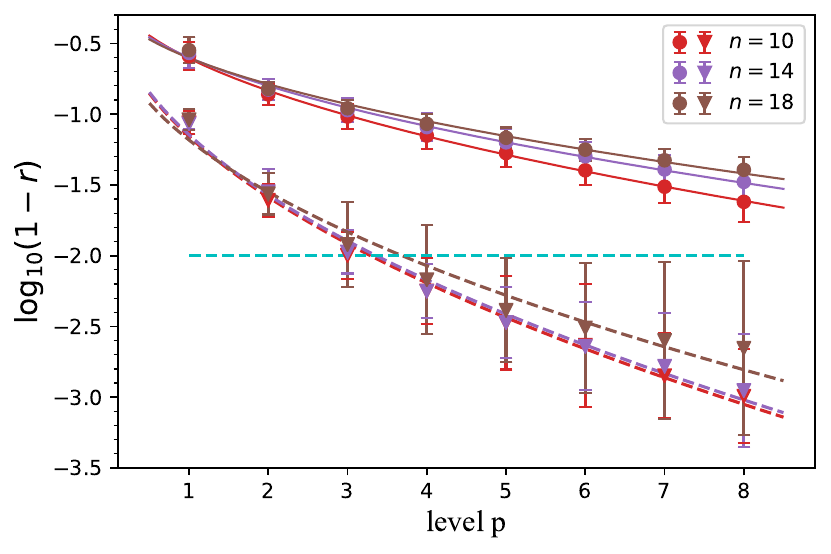}   
			\end{minipage}\label{fig:w3ra2}
		}
		{
			\begin{minipage}[h]{0.45\linewidth}
				\centering          
				\includegraphics[scale=0.52]{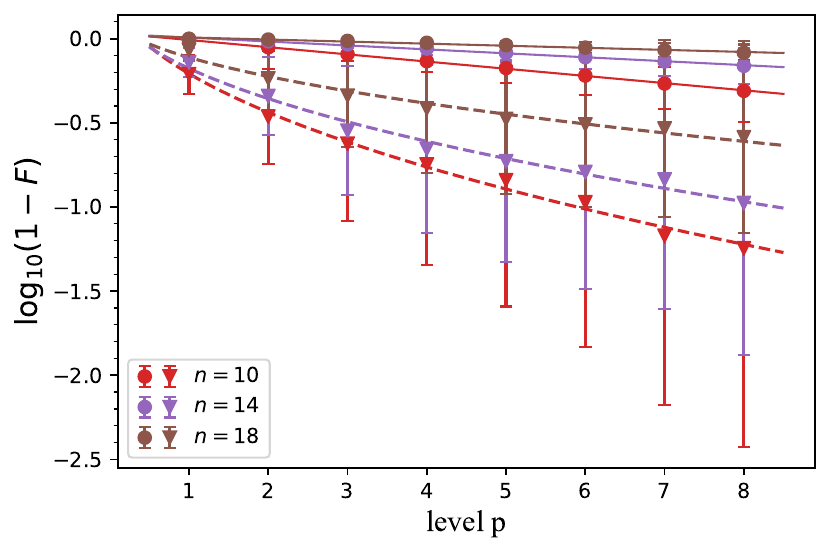}   
			\end{minipage}\label{fig:w3rb2}
		}\\
		{
			\begin{minipage}[h]{0.45\linewidth}
				\centering          
				\includegraphics[scale=0.52]{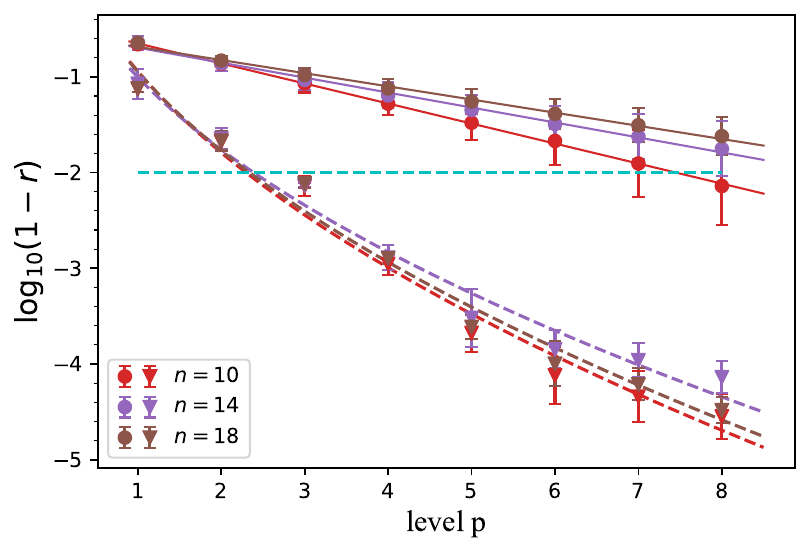}   
			\end{minipage}\label{fig:u3ra2}
		}
		{
			\begin{minipage}[h]{0.45\linewidth}
				\centering          
				\includegraphics[scale=0.52]{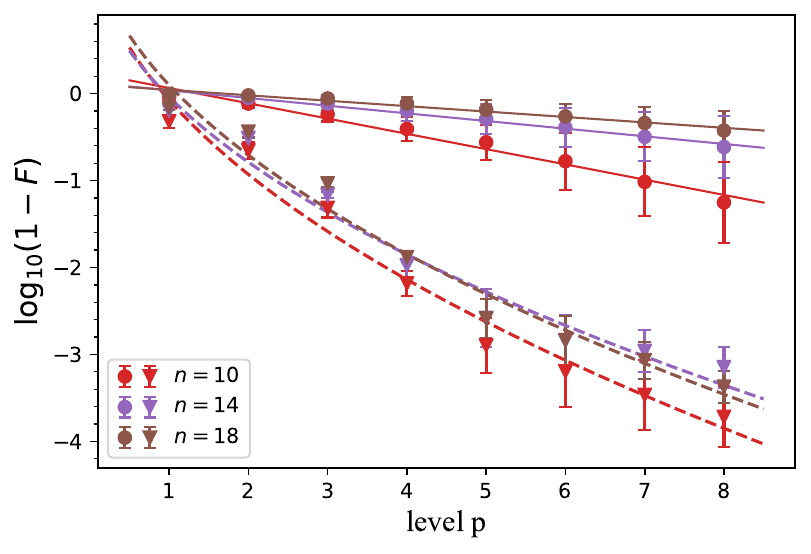}   
			\end{minipage}\label{fig:u3rb2}
		}
		
		\caption{Comparison of the accuracy ($\log_{10}(1-r)$) and infidelity ($\log_{10}(1-F)$) of results for QAOA (solid lines) and ab-QAOA (dashed lines) for $n=10,14,18$ for w3r graphs and u3r graphs. Top row is w3r graphs and bottom row is u3r graphs.  The horizontal dashed line represents $r^*=0.99$. Each point is an average over $40$ randomly chosen graphs. The fits are described in the main text. The error bars are standard deviations. As in Figs.~\ref{fig:w3r} and~\ref{fig:u3r} in the main text, the speedup of ab-QAOA over QAOA is clearly evident. The performances of both algorithms are slightly better in the unweighted case.  }\label{fig:wu3r2}
	\end{figure*}

	This appendix contains numerical results for both the w3r and u3r graphs with $n=10,14,18$. The same fitting functions are used as those in the main text and the fitting parameters are tabulated in Appendix~\ref{sec:fit}. The scaling analysis of these fitting functions will also be presented in Appendix~\ref{sec:fit}.

	The results for $n=18$ (brown triangles) in Figs.~B\ref{fig:u3ra2} and B\ref{fig:u3rb2} are below those for $n=14$ (purple triangles), also observed in Figs~\ref{fig:w3r} and ~\ref{fig:u3r}. From our results in Figs.~\ref{fig:u3ra} and B\ref{fig:u3ra2} and inspection of the local fields, it appears that there are some special  u3r graphs for which the ab-QAOA can find the solutions to the MaxCut problems using only a very shallow circuit depth. This creates the inversion with respect to $n$. In any case, one must keep in mind that the cost function is the energy, not the fidelity. A very low-lying excited state may have little overlap with the true ground state. We confirmed that the effect is mitigated by averaging over more graphs or by increasing $R$, the number of starting points, so we do not believe that there is anything very fundamental about it.

	\section{Fitting parameters}\label{sec:fit}

	In Table~\ref{tab:fit} we list the fitting parameters defined in the main text, which are computed using \textit{scipy.optimize.curve\_fit} function in python. The corresponding standard deviation errors $e_{p_0}$ and $e_c$ are also listed. On average, the fitting functions work better for the QAOA, where the fitting errors of $p_0$ for the u3r infidelity is a little bigger than those for the other cases. Overall, the fitting errors in ab-QAOA are bigger. However, what we care about most are the fits for the accuracy, which gives the estimated $p^*$ in the speedup, and the errors there are small.  
	
	We choose to fit all the points even though it might have been preferable  may be better to leave out the results in level $1$ when fitting the results of Fourier strategy, since the $R$ initial points are randomly generated instead of using the information from last level. If we did leave the $p=1$ points out, $e_{p_0}$ and $e_c$ in the QAOA would decrease slightly.  For the ab-QAOA, $e_{p_0}$ would decreased a small amount but $e_c$ would increase significantly, and there would be a noticeable deviation between the points and the fitting curves.  However, the main point about the fitting is the extrapolation of the QAOA data, so this does not affect any of our conclusions.  

\begin{figure*}[!htb]
		\centering    
		{
			\begin{minipage}[h]{0.45\linewidth}
				\centering          
				\includegraphics[scale=0.52]{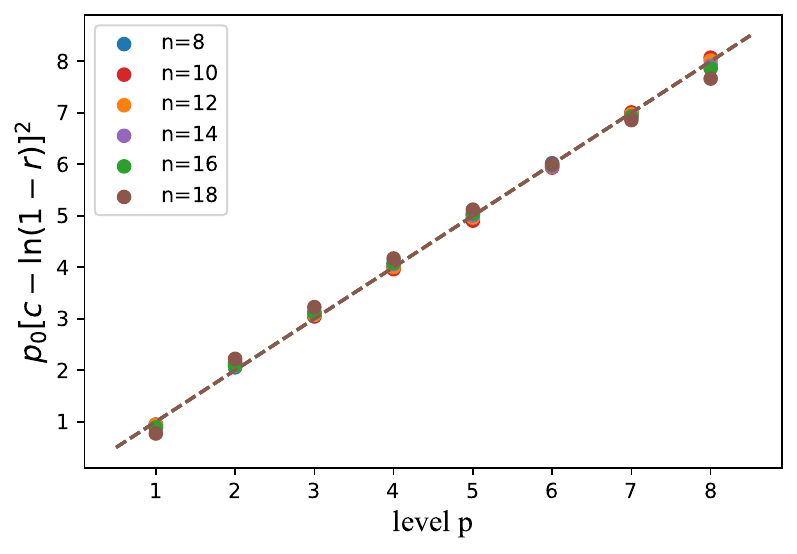}   
			\end{minipage}
		}
		{
			\begin{minipage}[h]{0.45\linewidth}
				\centering  
				\includegraphics[scale=0.52]{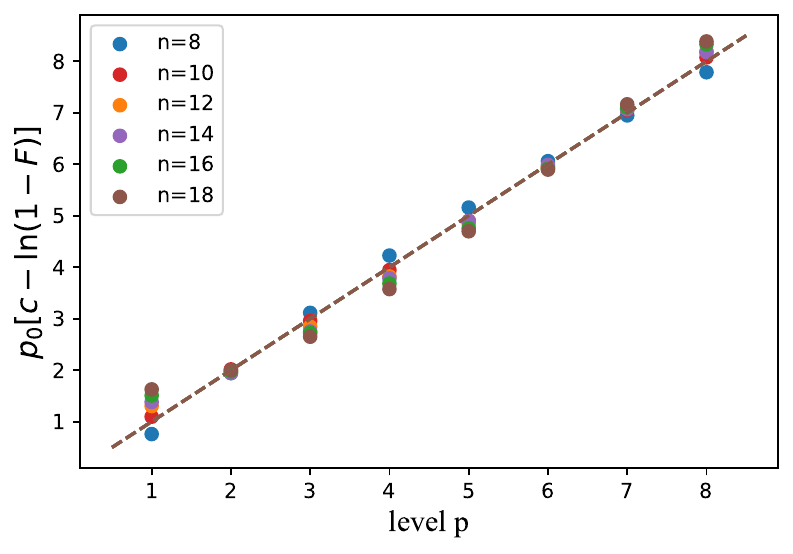}   
			\end{minipage}
		}\\
		
		{
			\begin{minipage}[h]{0.45\linewidth}
				\centering          
				\includegraphics[scale=0.52]{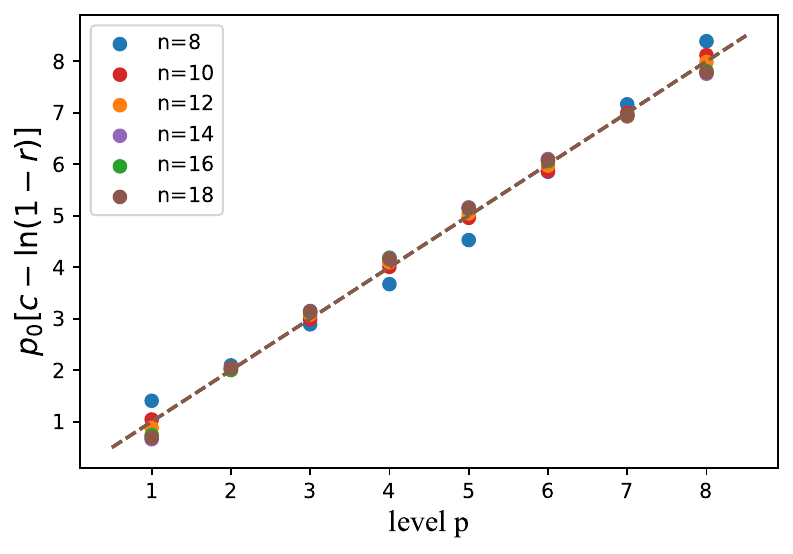}   
			\end{minipage}
		}
		{
			\begin{minipage}[h]{0.45\linewidth}
				\centering      
				\includegraphics[scale=0.52]{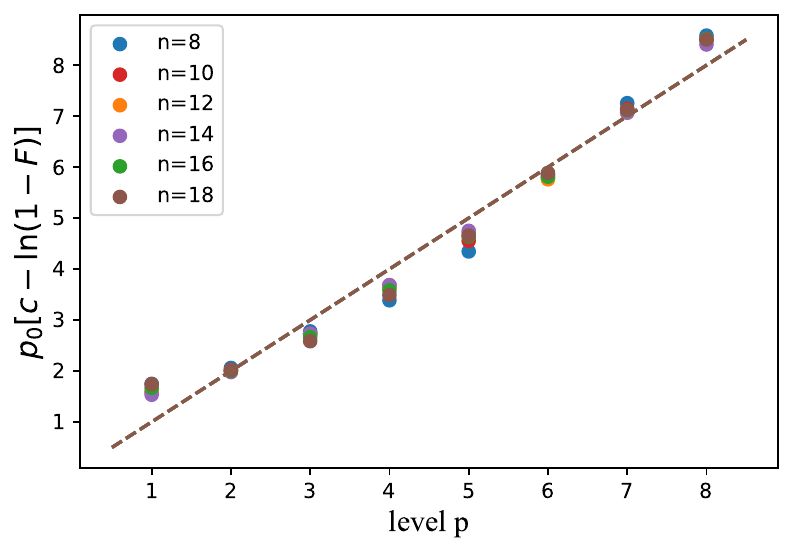}   
			\end{minipage}
		}\\
 
		{
			\begin{minipage}[h]{0.45\linewidth}
				\centering      
				\includegraphics[scale=0.52]{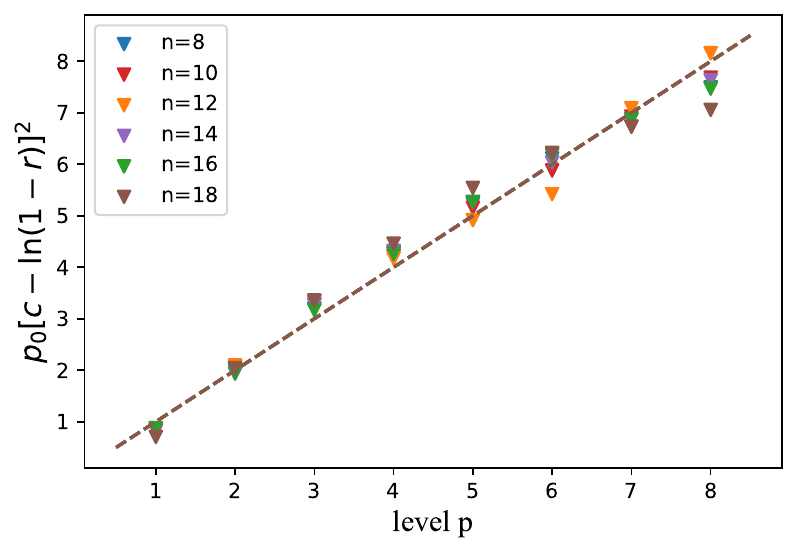}   
			\end{minipage}
		}
		{
			\begin{minipage}[h]{0.45\linewidth}
				\centering  
				\includegraphics[scale=0.52]{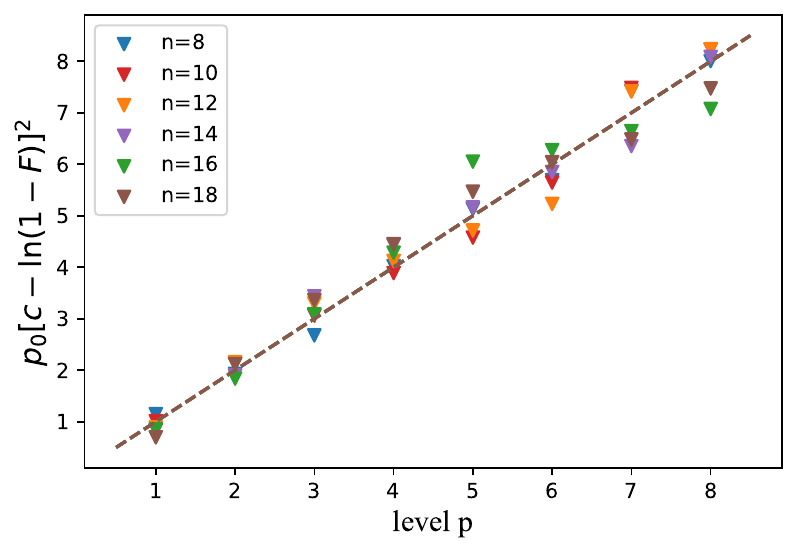}   
			\end{minipage}
		}\\
		
		{
			\begin{minipage}[h]{0.45\linewidth}
				\centering      
				\includegraphics[scale=0.52]{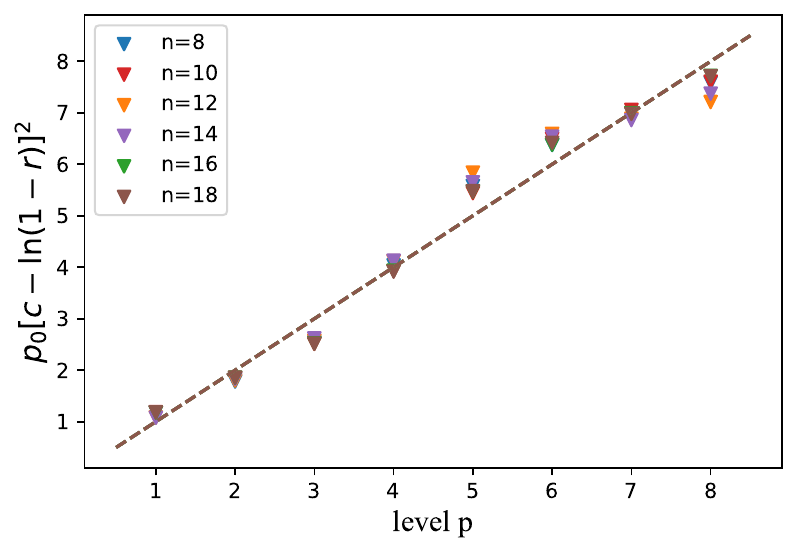}   
			\end{minipage}
		}
		{
			\begin{minipage}[h]{0.45\linewidth}
				\centering  
				\includegraphics[scale=0.52]{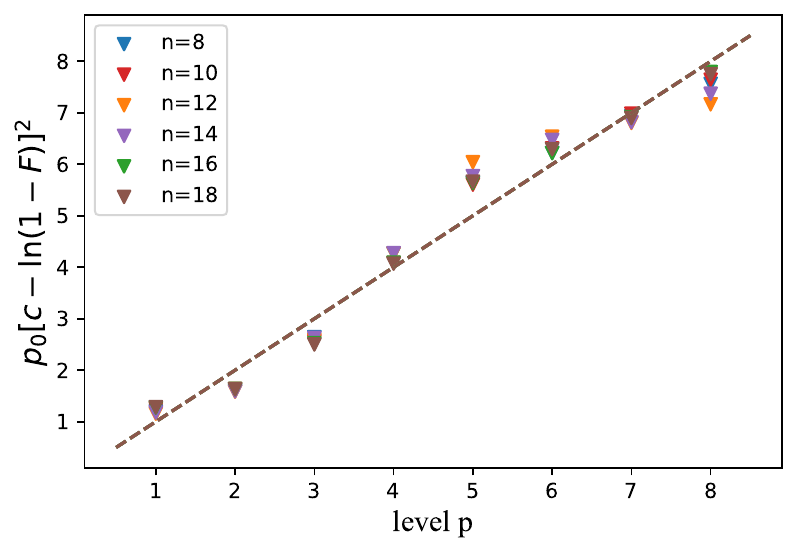}   
			\end{minipage}
		}
		\caption{Fits to the accuracy and infidelity curves for QAOA (top 2 rows) and ab-QAOA (bottom two rows). Rows 1 and 3 are for w3r graphs and rows 2 and 4 are for w3r graphs. The dashed lines in the the four subplots represent $p=p_0[c-\ln(1-r)]^2$ or $p=p_0[c-\ln(1-F)]^2$, which is equivalent to the fitting functions in the main text.}\label{fig:fit}
	\end{figure*}

	Using these fitting parameters and redefining the vertical axes of Figs.~\ref{fig:w3r} and \ref{fig:u3r} of the main text, we can collapse the graphs for the accuracy and infidelity onto straight lines, as shown in Fig.~\ref{fig:fit}.

	\begin{table*}[!htb]
		\centering
		\subtable
		{
			\begin{tabular}{|c|p{0.1\linewidth}<{\centering}|p{0.1\linewidth}<{\centering}|p{0.1\linewidth}<{\centering}|p{0.1\linewidth}<{\centering}|p{0.1\linewidth}<{\centering}|p{0.1\linewidth}<{\centering}|}
				\hline
				r for QAOA (w3r)
				{ }&$n=8$&$n=10$&$n=12$&$n=14$&$n=16$&$n=18$\\			
				\hline
				$p_0$&$0.4280$&$0.6223$&$0.7332$&$0.8023$&$0.9069$&$0.9443$\\
				\hline
				$e_{p_0}$&$0.0479$&$0.0514$&$0.0340$&$0.0592$&$0.0443$&$0.0925$\\
				\hline
				$c$&$0.1074$&$-0.1276$&$-0.2325$&$-0.2635$&$-0.3967$&$-0.3586$\\
				\hline	
				$e_{c}$&$0.0333$&$0.0430$&$0.0308$&$0.0562$&$0.0448$&$0.0953$\\
				\hline
			\end{tabular}
		}\\
		{        
			\begin{tabular}{|c|p{0.1\linewidth}<{\centering}|p{0.1\linewidth}<{\centering}|p{0.1\linewidth}<{\centering}|p{0.1\linewidth}<{\centering}|p{0.1\linewidth}<{\centering}|p{0.1\linewidth}<{\centering}|}
				\hline
				r for ab-QAOA (w3r)
				{ }&$n=8$&$n=10$&$n=12$&$n=14$&$n=16$&$n=18$\\
				\hline
				$p_0$&$0.1733$&$0.1758$&$0.1770$&$0.1796$&$0.1842$&$0.2392$\\
				\hline
				$e_{p_0}$&$0.4616$&$0.4285$&$0.4949$&$0.4619$&$0.4258$&$0.6381$\\
				\hline
				$c$&$-0.2451$&$-0.2796$&$-0.2611$&$-0.2788$&$-0.2483$&$-0.6791$\\
				\hline		
				$e_{c}$&$0.2038$&$0.1906$&$0.2209$&$0.2076$&$0.1939$&$0.3310$\\
				\hline
			\end{tabular}
		}\\
		\subtable
		{        
			\begin{tabular}{|c|p{0.1\linewidth}<{\centering}|p{0.1\linewidth}<{\centering}|p{0.1\linewidth}<{\centering}|p{0.1\linewidth}<{\centering}|p{0.1\linewidth}<{\centering}|p{0.1\linewidth}<{\centering}|}
				\hline
				F for QAOA (w3r)
				{ }&$n=8$&$n=10$&$n=12$&$n=14$&$n=16$&$n=18$\\
				\hline
				$p_0$&$8.4398$&$10.1726$&$12.8145$&$18.5465$&$29.5196$&$35.3103$\\
				\hline
				$e_{p_0}$&$0.0034$&$0.0012$&$0.0023$&$0.0019$&$0.0017$&$0.0018$\\
				\hline
				$c$&$0.0340$&$0.0805$&$0.0881$&$0.0695$&$0.0485$&$0.0453$\\
				\hline	
				$e_{c}$&$0.0170$&$0.0060$&$0.0116$&$ 0.0095$&$0.0085$&$0.0089$\\
				\hline
			\end{tabular}
		}\\
		\subtable
		{        
			\begin{tabular}{|c|p{0.1\linewidth}<{\centering}|p{0.1\linewidth}<{\centering}|p{0.1\linewidth}<{\centering}|p{0.1\linewidth}<{\centering}|p{0.1\linewidth}<{\centering}|p{0.1\linewidth}<{\centering}|}
				\hline
				F for ab-QAOA (w3r)
				{ }&$n=8$&$n=10$&$n=12$&$n=14$&$n=16$&$n=18$\\
				\hline
				$p_0$&$0.5021$&$0.6129$&$0.7549$&$1.0054$&$1.5762$&$2.5342$\\
				\hline
				$e_{p_0}$&$0.1585$&$0.1335$&$0.1490$&$0.1112$&$0.0992$&$0.0551$\\
				\hline
				$c$&$1.0155$&$0.7984$&$0.6930$&$0.5899$&$0.5140$&$0.3716$\\
				\hline	
				$e_{c}$&$0.1192$&$0.1109$&$0.1373$&$0.1188$&$0.1321$&$0.0930$\\
				\hline
			\end{tabular}
		}\\
	
		\subtable
		{        
			\begin{tabular}{|c|p{0.1\linewidth}<{\centering}|p{0.1\linewidth}<{\centering}|p{0.1\linewidth}<{\centering}|p{0.1\linewidth}<{\centering}|p{0.1\linewidth}<{\centering}|p{0.1\linewidth}<{\centering}|}
				\hline
				r for QAOA (u3r)
				{ }&$n=8$&$n=10$&$n=12$&$n=14$&$n=16$&$n=18$\\
				\hline
				$p_0$&$1.3023$&$2.0753$&$2.4677$&$2.7788$&$3.0333$&$3.1562$\\
				\hline
				$e_{p_0}$&$0.0694$&$0.0246$&$0.0304$&$0.0917$&$0.0800$&$0.0937$\\
				\hline
				$c$&$-0.5508$&$-1.0162$&$-1.1651$&$-1.2412$&$-1.2671$&$-1.2632$\\
				\hline	
				$e_{c}$&$0.2069$&$0.0288$&$0.0252$&$0.0600$&$0.0439$&$0.0475$\\
				\hline
			\end{tabular}
		}\\
		\subtable
		{        
			\begin{tabular}{|c|p{0.1\linewidth}<{\centering}|p{0.1\linewidth}<{\centering}|p{0.1\linewidth}<{\centering}|p{0.1\linewidth}<{\centering}|p{0.1\linewidth}<{\centering}|p{0.1\linewidth}<{\centering}|}
				\hline
				r for ab-QAOA (u3r)
				{ }&$n=8$&$n=10$&$n=12$&$n=14$&$n=16$&$n=18$\\
				\hline
				$p_0$&$0.0481$&$0.0448$&$0.0541$&$ 0.0565$&$0.0474$&$0.0478 $\\
				\hline
				$e_{p_0}$&$0.0058$&$0.0053$&$0.0080$&$0.0070$&$0.0052$&$0.0054$\\
				\hline
				$c$&$2.4151$&$2.5535$&$1.9384$&$1.8957$&$2.3950$&$2.3987$\\
				\hline
				$e_{c}$&$0.5815$&$0.5896$&$0.6777$&$0.5551$&$0.5354$&$0.5466$\\
				\hline
			\end{tabular}
		}\\
		{        
			\begin{tabular}{|c|p{0.1\linewidth}<{\centering}|p{0.1\linewidth}<{\centering}|p{0.1\linewidth}<{\centering}|p{0.1\linewidth}<{\centering}|p{0.1\linewidth}<{\centering}|p{0.1\linewidth}<{\centering}|}
				\hline
				F for QAOA (u3r)
				{ }&$n=8$&$n=10$&$n=12$&$n=14$&$n=16$&$n=18$\\
				\hline
				$p_0$&$1.2631$&$ 2.4719$&$3.3318$&$4.9315$&$5.8239$&$6.9639$\\
				\hline
				$e_{p_0}$&$0.1080$&$0.1576$&$0.2050$&$0.2623$&$0.3985$&$0.5160$\\
				\hline
				$c$&$1.1888$&$0.5537$&$0.4275$&$0.2897$&$0.2734$&$0.2436$\\
				\hline		
				$e_{c}$&$0.3418$&$0.1302$&$0.0932$&$0.0545$&$0.0593$&$0.0537$\\
				\hline
			\end{tabular}
		}\\
		\subtable
		{
			\begin{tabular}{|c|p{0.1\linewidth}<{\centering}|p{0.1\linewidth}<{\centering}|p{0.1\linewidth}<{\centering}|p{0.1\linewidth}<{\centering}|p{0.1\linewidth}<{\centering}|p{0.1\linewidth}<{\centering}|}
				\hline
				F for ab-QAOA (u3r)
				{ }&$n=8$&$n=10$&$n=12$&$n=14$&$n=16$&$n=18$\\
				\hline
				$p_0$&$0.0496$&$0.0442$&$0.0519$&$0.0574$&$0.0490$&$0.0499$\\
				\hline
				$e_{p_0}$&$0.0067$&$ 0.0060$&$0.0093$&$0.0088$&$0.0067$&$0.0072$\\
				\hline
				$c$&$4.1400$&$4.5860$&$4.1717$&$4.0884$&$4.6724$&$4.7042$\\
				\hline
				$e_{c}$&$0.6389$&$0.6792$&$0.8383$&$0.6831$&$0.6599$&$0.6857$\\
				\hline
			\end{tabular}
		}
		\caption{Fitting parameters $p_0$ and $c$ of QAOA and ab-QAOA for w3r graphs and u3r graphs. $e_{p_0}$ and $e_c$ represent the standard deviation errors. Top left entry in each table specifies accuracy (r) or infidelity (F), algorithm, and graph type.}\label{tab:fit}
	\end{table*}	
    \newpage
	
	\section{Speedup parameter $p^*$}\label{sec:p*}

	In Table~\ref{tab:p}, we list $p^*$ for the calculation of the speedup $S(n)$ shown in Fig.~\ref{fig:speedup} in the main text.  For the QAOA, $p^*$ is obtained from the fitting function. For the ab-QAOA, $p^*$ is obtained directly from the numerical simulation. For clarity of the speedup, all $p^{*}$ are rounded to integers.
	
	\begin{table}[H]
		\centering
		\subtable
		{
			\begin{tabular}{|c|c|c|c|c|c|c|}
				\hline
				w3r
				{ }&$n=8$&$n=10$&$n=12$&$n=14$&$n=16$&$n=18$\\
				\hline
				standard QAOA&$10$&$12$&$14$&$15$&$16$&$17$\\
				\hline
				ab-QAOA&$3$&$3$&$3$&$3$&$3$&$3$\\
				\hline			
			\end{tabular}
		}\\
		\subtable
		{        
			\begin{tabular}{|c|c|c|c|c|c|c|}
				\hline
				u3r
				{ }&$n=8$&$n=10$&$n=12$&$n=14$&$n=16$&$n=18$\\
				\hline
				standard QAOA&$5$&$7$&$8$&$9$&$10$&$11$\\
				\hline
				ab-QAOA&$3$&$3$&$3$&$3$&$3$&$3$\\
				\hline			
			\end{tabular}
		}
		\caption{$p^*$ in speedup for w3r and u3r graphs}\label{tab:p}
	\end{table}
    
	\bibliography{mainref}
	
\end{document}